\documentclass[preprint,12pt,authorhyear]{elsarticle}
\usepackage{geometry}
\usepackage{mathtools}
\usepackage[utf8]{inputenc}
\usepackage[T1]{fontenc}
\usepackage{verbatim}
\usepackage{amsmath,amssymb,amsfonts}	
\usepackage{graphicx}
\usepackage{color}
\usepackage{xcolor}
\usepackage{physics}
\usepackage{booktabs}
\usepackage[normalem]{ulem}
\usepackage{appendix}
\usepackage{slashed}
\usepackage{tikz}
\usetikzlibrary{angles,quotes}
\usetikzlibrary{decorations.pathmorphing,decorations.markings,arrows.meta}
\usepackage[colorlinks=true,citecolor=blue,linkcolor=purple,urlcolor=purple]{hyperref}
\usepackage{array}

\begin{document}
\begin{frontmatter}
\title{Z and Higgs boson decays with doubly-charged scalars at one-loop: current constraints, future sensitivities, and application to lepton-triality models.}

\author[Sydney]{Gabriela Lichtenstein}
\ead{gabriela.lichtenstein@alumni.usp.br}

\author[Sydney]{Michael A.~Schmidt}
\ead{m.schmidt@unsw.edu.au}

\author[Monash]{German Valencia}
\ead{german.valencia@monash.edu}

\author[Melbourne]{Raymond R.~Volkas}
\ead{raymondv@unimelb.edu.au}
\affiliation[Sydney]{
organization = {Sydney Consortium for Particle Physics and Cosmology, School of Physics, The University of New South Wales},
city={Sydney}, 
state={New South Wales},
postcode={2052},
country={Australia}
}

\affiliation[Monash]{
organization={School of Physics and Astronomy, Monash University}, 
addressline={Wellington Road}, 
city={Clayton}, 
state={Victoria},
postcode={3800},
country={Australia}}

\affiliation[Melbourne]{
organization={ARC Centre of Excellence for Dark Matter Particle Physics, School of Physics, The University of Melbourne}, 
state={Victoria},
postcode={3010}, 
country={Australia}}

\begin{abstract}
We analyse the $Z$ and Higgs boson decays $Z\to \ell^+ \ell^- $ ($\ell =  e, \mu,\tau$), $H\to \gamma\gamma$ and $H\to Z\gamma$ that are induced at one-loop level in models with a doubly-charged isosinglet scalar. After discussing current constraints, we derive the parameter space that will be probed by the HL-LHC and the possible future colliders the ILC, CEPC and FCC. We then apply those constraints to lepton triality models which are based on a discrete $Z_3$ family symmetry and were recently studied in the context of charged-lepton flavour-violating processes at Belle II and the proposed $\mu^+ \mu^+$ and $\mu^+ e^-$ collider known as $\mu$TRISTAN. We find that the future constraints that can be imposed by $Z \to \ell^+ \ell^-$ on the lepton flavour conserving couplings of the triality models reduce the viable parameter space to probe lepton flavour violating processes. The constraints from Higgs boson decays are the first on the Higgs portal sector of the triality models.    
\end{abstract}

\end{frontmatter}

\section{Introduction}

The phenomenology of doubly-charged scalars has been a topic of on-going interest~\cite{deBlas:2014mba,Babu:2016rcr,ATLAS:2017xqs,Crivellin:2018ahj,Li:2018cod,Li:2019xvv,deMelo:2019asm,Fuks:2019clu,Anisha:2021jlz,Anisha:2021hgc,ATLAS:2022pbd,Ruiz:2022sct,Xu:2023ene}. In this paper, we study isosinglet doubly-charged scalars that couple to right-handed charged-lepton bilinears and induce contributions at one-loop level to the Higgs boson decays $H \to \gamma \gamma,\, Z\gamma$ and the $Z$ boson decays $Z \to \ell^+ \ell^-$ ($\ell = e, \mu, \tau$). As well as the general model of this kind, we also apply our results to specific realisations employing lepton triality (see below).

After computing the most up-to-date bounds from existing measurements of these processes, we also determine for the first time the parameter space reach of the High-Luminosity Large Hadron Collider (HL-LHC) ~\cite{Cepeda:2019klc}, an International Linear Collider (ILC) ~\cite{ILC:2013jhg, Bambade:2019fyw, ILCInternationalDevelopmentTeam:2022izu}, a Circular Electron Positron Collider (CEPC) ~\cite{CEPCStudyGroup:2018ghi,Gao:2022lew, CEPCPhysicsStudyGroup:2022uwl}, and a Future Circular Collider in both $e^+ e^-$ mode (FCC-ee) and through an integrated $ee/eh/hh$ program (FCC-int)~\cite{FCC:2018byv, Bernardi:2022hny}.

Our analysis goes beyond the recent analyses~\cite{deBlas:2014mba,Anisha:2021hgc} of a doubly-charged scalar which have been carried out as part of a more general study of single scalar particle extensions. Reference \cite{deBlas:2014mba} did not include loop-level matching of the single scalar extensions onto SM effective field theory (SMEFT) and thus does not include all contributions to the leptonic $Z$-boson couplings. Reference~\cite{Anisha:2021hgc} performed the matching onto SMEFT at the one-loop level and carried out a general study within SMEFT, but did not translate the general results in SMEFT back to the model parameters of the doubly-charged scalar. We further use the latest experimental results for Higgs decays and projections for future colliders in our analysis.

The general results are used to determine future experimental sensitivities and current constraints on specific isosinglet doubly-charged scalar models that utilise the discrete flavour symmetry of lepton triality.\footnote{Lepton triality may appear as an approximate symmetry in more complete  lepton flavour discrete symmetry models; see e.g.~Refs.~\cite{Altarelli:2005yx,He:2006dk,deAdelhartToorop:2010jxh,deAdelhartToorop:2010nki,Cao:2011df,Holthausen:2012wz,Pascoli:2016wlt,Muramatsu:2016bda}.} These models are of interest because they serve as benchmarks for the phenomenological study of charged-lepton flavour-violating (CLFV) processes~\cite{Bigaran:2022giz,Lichtenstein:2023iut}.  They feature doubly-charged scalars with family-dependent Yukawa couplings to charged fermions that are dictated by lepton triality. The main motivation for studying these particular models is to provide simple exemplar theories where the dominant beyond Standard Model (BSM) effects are flavour-changing processes involving the $\tau$ lepton, in particular the decays $\tau^\pm \to \mu^\pm \mu^\pm e^\mp$ and $\tau^\pm \to e^\pm e^\pm \mu^\mp$ that will be probed with impressive sensitivity by Belle II~\cite{Bigaran:2022giz}. Another phenomenological analysis~\cite{Lichtenstein:2023iut} compared the Belle II reach from $\tau$ decays with complementary bounds from charged-lepton scattering processes that could eventually be obtained using the proposed $\mu^+ \mu^+$ and $\mu^+ e^-$ collider facility known as $\mu$TRISTAN~\cite{Hamada:2022mua}.

The remainder of this paper is structured as follows. Section~\ref{sec:doubly-charged} introduces the doubly-charged scalar extension of the SM, reviews the isosinglet triality models of Ref.~\cite{Bigaran:2022giz} and summarises the currently known experimental bounds.  Section~\ref{sec:constraints} then derives constraints from the one-loop contributions to
$Z \to \ell^+ \ell^-$, $H \to \gamma \gamma$ and $H \to Z \gamma$ based on existing data, and also determines the parameter-space reach of HL-LHC, ILC and FCC. Section~\ref{sec:conc} is a conclusion. Some technical details are provided in the Appendices.

\section{Doubly-charged scalar models}
\label{sec:doubly-charged}

\subsection{General model}
\label{sec:general-model}

We consider a colourless doubly-charged isosinglet scalar $k$ with Yukawa interactions
\begin{equation}\label{eq:Lk}
    \mathcal{L}_k = \frac12 y_{ij} \overline{(e_{iR})^c} e_{jR} k + \mathrm{h.c.}.
\end{equation}
The Yukawa coupling matrix $y$ is symmetric in the flavour indices, $y_{ij}=y_{ji}$. We denote the mass of the exotic scalar by $m_k$. 
The scalar $k$ also couples to the SM Higgs doublet $\phi$ via the quartic Higgs portal interaction, and of course also to the photon $A^\mu$ and the $Z$-boson $Z^\mu$. 
The relevant non-Yukawa interaction terms are 
\begin{equation}
 \mathcal{L} \supset |D_\mu k|^2  -m_{k}^2 k^\dagger k
 - \frac{\lambda_k}{2} (k^\dagger k)^2
 - \kappa_{\phi}\, \left(\phi^\dagger \phi -\frac{v^2}{2}\right) \, k^\dagger k 
 - \frac{\lambda}{2} \left(\phi^\dagger \phi - \frac{v^2}{2}\right)^2
\end{equation}
where $D_\mu= \partial_\mu + i\, 2 e\, ( A_\mu - \tan\theta_W Z_\mu)$ is the covariant derivative, with $\theta_W$ being the weak mixing angle, $e$ the electromagnetic coupling constant, $\lambda$ the quartic SM Higgs interaction, $\lambda_k$ the quartic $k$ interaction, $v=(\sqrt{2}G_F)^{-1/2}\simeq 246$ GeV the SM Higgs vacuum expectation value, and $\kappa_\phi$ the Higgs portal coupling constant between the SM Higgs doublet and $k$. In unitary gauge after electroweak symmetry breaking $\phi^\dagger\phi = (v+H)^2/2$, where $H$ denotes the Higgs field. 
Perturbative unitarity constrains the Yukawa couplings to be smaller than $\sqrt{4\pi}$ and the scalar couplings $|\lambda_{k}|\leq 4\pi$ and $|\kappa_\phi|\leq 8\pi$. The stability of the scalar potential requires $\lambda, \lambda_k > 0$ and $\kappa_\phi > - \sqrt{\lambda\lambda_k} > -\sqrt{4\pi} m_H/v \simeq -1.8$; see e.g.~\cite{Kannike:2012pe}, where we used the perturbative unitarity bound for $\lambda_k$ in the last inequality.\footnote{The study of vacuum stability dependence on the renormalisation scale depends on unknown ultraviolet physics which we expect to be present and thus it is beyond the scope of this work.}

The specific class of doubly-charged isosinglet scalar models utilising lepton triality is briefly reviewed next.

\subsection{Brief review of the lepton triality models}
\label{sec:triality}

The $Z_3$ flavour symmetry of lepton triality acts solely on leptons via the transformations
\begin{equation}
    L \to \omega^T L,\qquad e_R \to \omega^T e_R
\end{equation}
where $T=1,2,3$ for the first, second and third families, respectively, and $\omega = e^{2\pi i/3}$ is a cube root of unity. This means that $e$ and $\nu_e$ transform multiplicatively via $\omega$, $\mu$ and $\nu_\mu$ transform via $\omega^2$, while $\tau$ and $\nu_\tau$ are singlets. The charged-lepton Yukawa and mass matrices are forced to be diagonal. See Ref.~\cite{Bigaran:2022giz} for a brief discussion of neutrino mass generation within the lepton triality models.

The models we consider all contain a single colourless, isosinglet, doubly-charged  scalar. There are three models, each containing one of the positive doubly-charged scalars $k_1$, $k_2$ or $k_3$, with 
\begin{equation}
k_1 \to \omega k_1,\quad k_2\to \omega^2 k_2\quad \text{and}\quad k_3 \to k_3 
\end{equation}
under triality. We refer to these as the $T=1$, $T=2$ and $T=3$ models, respectively.\footnote{The isotriplet models introduced in Ref.~\cite{Bigaran:2022giz} will not be considered in this paper, but would be interesting to examine in future work.} 
We denote the masses of the exotic scalars by $m_{k_1}$, $m_{k_2}$ and $m_{k_3}$. 

Triality constrains the allowed Yukawa interactions from Eq.~\eqref{eq:Lk}: $k_1$ is restricted to $y_{32}= y_{23} \equiv f_1$ and $y_{11} \equiv f_2$ with all others vanishing; $k_2$ has $y_{31}= y_{13} \equiv g_1$ and $y_{22} \equiv g_2$ with the rest being zero; and $k_3$ has $y_{21}= y_{12} \equiv h_1$ and $y_{33} \equiv h_2$ only.
The coupling constants $f_{1,2}$, $g_{1,2}$ and $h_{1,2}$ may be taken to be real and non-negative without loss of generality.

\subsection{Existing constraints}
\label{sec:existing-constraints}

We now very briefly summarise some existing experimental constraints -- see Refs.~\cite{Bigaran:2022giz,Lichtenstein:2023iut} for further discussion, especially in the context of the lepton triality models.

There are two analyses using LHC data of direct searches for doubly-charged scalars decaying to same-sign charged lepton pairs, including flavour-violating modes. First, ATLAS \cite{ATLAS:2017xqs} conducted a search with an integrated luminosity of 36 fb$^{-1}$ of data for the pair production of doubly-charged scalar singlets decaying to $\ell^\pm \ell^\pm$. Not seeing a signal, they derived direct lower bounds on the scalar masses. The precise bounds depend on assumptions about the branching ratios, but they are in the approximate range of $0.55-0.75$ TeV for typical benchmark choices. Second, as discussed previously in Ref.~\cite{Lichtenstein:2023iut}, the latest ATLAS searches~\cite{ATLAS:2022pbd} using an integrated luminosity of 139 fb$^{-1}$ obtained an improved lower bound of $0.9$ TeV on the doubly-charged scalar mass in the Zee-Babu model ~\cite{Zee:1985id,Babu:1988ki} for the special case of universal lepton couplings. This limit applies to the doubly-charged scalars in triality models if they decay to same-sign leptons with equal probability for different flavours, which is not the general situation. In the absence of precise bounds expressed as functions of different branching ratios for the 139 fb$^{-1}$ analysis, we show two \textit{indicative} bounds in the later plots, labelled by the integrated luminosities 36 fb$^{-1}$ and 139 fb$^{-1}$.

Reference~\cite{Bigaran:2022giz} analysed existing Belle constraints on the CLFV decay modes of the $\tau$, producing the bounds
\begin{equation}
    \sqrt{f_1 f_2} \lesssim 0.17 \frac{m_{k_1}}{\mathrm{TeV}}\quad \text{and}\quad
    \sqrt{g_1 g_2} \lesssim 0.17 \frac{m_{k_2}}{\mathrm{TeV}},
    \label{Eq:UpperBound}
\end{equation}
with the future sensitivity of Belle II~\cite{Bigaran:2022giz} determined to be 
\begin{equation}
    \sqrt{f_1 f_2} \lesssim 0.06 \frac{m_{k_1}}{\mathrm{TeV}}\quad \text{and}\quad
    \sqrt{g_1 g_2} \lesssim 0.06 \frac{m_{k_2}}{\mathrm{TeV}}.
\end{equation}
Note that the scalar $k_3$ cannot be probed by $\tau$ decays at tree level. See Ref.~\cite{Lichtenstein:2023iut} for the potential sensitivity of $\mu$TRISTAN.

We also consider constraints from leptonic anomalous magnetic dipole moments.  The electromagnetic current of a lepton of mass $m_\ell$ is parametrised in terms of three form factors $F_i$ for a real on-shell photon. In the CP conserving case, they reduce to the Dirac form factor $F_1$ and the Pauli form factor $F_2$, see e.g.~the discussion in~\cite{Nowakowski:2004cv,Schwartz:2014sze},
\begin{equation}
    \braket{\ell(\mathbf{p}_2)|j^\mu_{\rm em}(0)}{\ell(\mathbf{p}_1)} = \bar u(\mathbf{p}_2)\left[ F_1(q^2) \gamma^\mu + F_2(q^2) \frac{i\sigma^{\mu\nu}}{2m_\ell}q_\nu 
    \right] u(\mathbf{p}_1)
    \qquad\mathrm{with}\qquad p_1 = p_2+q\;.
\end{equation}
At $q^2=0$ the form factors are identified with the electric charge $F_1(0) = Q_\ell e$ and the anomalous magnetic moment $a_\ell = F_2(0)/F_1(0)$. We calculated the contribution of the doubly charged scalar to the anomalous magnetic moment using the general formulae in~\cite{Lavoura:2003xp} and independently using \texttt{FeynCalc}~\cite{Mertig:1990an,Shtabovenko:2016sxi,Shtabovenko:2020gxv,Shtabovenko:2023idz}, \texttt{FeynHelpers}~\cite{Shtabovenko:2016whf} and \texttt{Package-X}~\cite{Patel:2015tea,Patel:2016fam}. The leading order contribution to the anomalous magnetic moment in an expansion in $m_\ell/m_k$ is given by
\begin{equation}\label{eq:AMMtheory}
    \Delta a_\ell =  - \frac{(y^\dagger y)_{\ell\ell}}{24\pi^2} \frac{m_\ell^2}{m_k^2}
\end{equation}
which agrees in magnitude with \cite{Li:2019xvv}, but corrects the overall sign. It is also consistent with the result in~\cite{Xu:2023ene}. 

There are two independent determinations of the electron anomalous magnetic moment, one using cesium~\cite{Parker:2018vye} and the other rubidium~\cite{Morel:2020dww} which result in the constraints 
\begin{align}
    \Delta a_e^{\rm  Cs} & = (-8.8\pm 3.6) \times 10^{-13}
     &
     \Rightarrow \qquad 
     (y^\dagger y)_{ee}^{1/2} &< 34 [38]\,  \frac{m_k}{\mathrm{TeV}}\;, 
    \\
    \Delta a_e^{\rm Rb} & = (4.8 \pm 3.0) \times 10^{-13}
     &
     \Rightarrow \qquad 
     (y^\dagger y)_{ee}^{1/2} &< [10] \,\frac{m_k}{\mathrm{TeV}} 
 \end{align}
at $1 [2]\sigma$. Due to the difference in sign of the predicted $\Delta a_e$ in Eq.~\eqref{eq:AMMtheory} and the central value of the rubidium measurement, the doubly charged isosinglet scalar is only compatible with the $\Delta a_e$ measurement in rubidium at $2\sigma$.

For the anomalous magnetic moment of the muon, we conservatively compare the experimental results~\cite{Muong-2:2006rrc,Muong-2:2023cdq} with the theory prediction in the white paper~\cite{Aoyama:2020ynm}, where the hadronic vacuum polarization has been replaced by the recent lattice result of the BMW collaboration~\cite{Borsanyi:2020mff,Boccaletti:2024guq} following~\cite{Athron:2024rir}. Demanding the deviation to be less than $1 [2]\sigma$ we find
 \begin{align}
    \Delta a_\mu &= (4.0\pm4.4)\times 10^{-10} &
     \Rightarrow \qquad 
     (y^\dagger y)_{\mu\mu}^{1/2} &< 0.92 [3.2] \,\frac{m_k}{\mathrm{TeV}} \;.
 \end{align}

\mathversion{bold}
\section{One-loop-induced processes: constraints and prospects }
\mathversion{normal}
\label{sec:constraints}

We now turn our attention to the main section of this paper: an analysis of constraints arising from $1$-loop contributions to $Z \to \ell^+ \ell^-$, $H\to \gamma \gamma$ and $H \to Z \gamma$ from existing data, and the determination of parameter-space reach at future colliders. We focus on flavour conserving $Z$ boson decays but it is straightforward to generalise our results to flavour violating $Z$ boson decays. Note that in the case of the triality models, that symmetry forbids CLFV violating $Z \to \ell^+ \ell^{\prime-}$ decays.

\subsection{\texorpdfstring{$Z \to \ell^+ \ell^- $}{Z->l+l-}}
\label{sub:Zll}

\begin{figure}[tb!]
    \centering
     \begin{tikzpicture}[scale =0.4,
      ->-/.style={decoration={markings,mark=at position 0.6 with {\arrow{Stealth}}},postaction={decorate}},
      ]
  \filldraw[black] (0,-1) node{$Z$};
 \draw [-,black, decorate,decoration=snake] (-2,0) -- (2,0);
 
 \draw[black, thick, ->-] (2,0) -- (4,-2);
 \filldraw[black] (2.7,-1.8) node{$\ell'$};
 \draw[black, thick,->-] (6,-3) -- (4,-2);
 \filldraw[black] (7,-3.5) node{$\ell^+$};
 
 \draw[black, dashed,->-](4,2) --  (4,-2);
\filldraw[black] (5,0)  node {$k$};

 \draw[black, thick,->-] (4,2) -- (2,0);
 \filldraw[black] (2.7,1.8) node{$\ell'$};
 \draw[black, thick, ->-] (4,2) -- (6,3);
 \filldraw[black] (7,3.5) node{$\ell^-$};

 \end{tikzpicture}%
\hspace{1cm}
  \begin{tikzpicture}[scale =0.4,
  ->-/.style={decoration={markings,mark=at position 0.6 with {\arrow{Stealth}}},postaction={decorate}},
  ]
  \filldraw[black] (0,-1) node{$Z$};
 \draw [-,black, decorate,decoration=snake] (-2,0) -- (2,0);
 
 \draw[black, dashed,->-] (2,0) -- (4,-2);
 \filldraw[black] (2.7,-1.8) node{$k$};
 \draw[black, thick,->-] (6,-3) -- (4,-2);
 \filldraw[black] (7,-3.5) node{$\ell^+$};
 
 \draw[black, thick, ->-] (4,2) -- (4,-2);
\filldraw[black] (5,0)  node{$\ell'$};

 \draw[black, dashed,->-] (4,2) -- (2,0);
 \filldraw[black] (2.7,1.8) node{$k$};
 \draw[black, thick, ->-] (4,2) -- (6,3);
 \filldraw[black] (7,3.5) node{$\ell^-$};

 \end{tikzpicture}
    
    \caption{The doubly-charged scalar vertex corrections to $Z\to \ell^+ \ell^-$. There are also self-energy corrections that we do not show. 
    All corrections have been incorporated in the final result. The virtual lepton $\ell'$ may be of a different flavour from the final state leptons $\ell$.}
    \label{f:Zll-diagrams}
\end{figure}
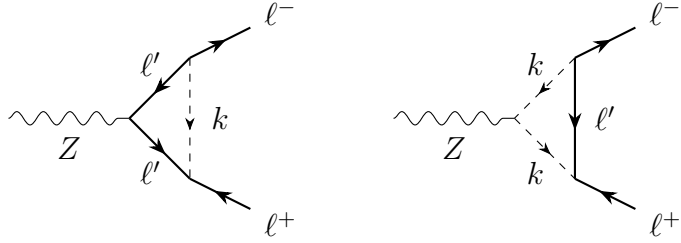

The Yukawa couplings $y_{ij}$ in Eq.~\eqref{eq:Lk} modify the couplings of the $Z$ boson to charged leptons at one loop level.
Since the flavour conserving $Z$ boson couplings have been measured very precisely at LEP and SLD, we expect that stringent constraints may arise. We show the one-loop vertex corrections due to $k_i$ in Fig.~\ref{f:Zll-diagrams}.

To compute these corrections we write the tree plus one-loop vertex factor as
\begin{equation}
\Gamma^\ell_\mu = -~2^{1/4}G_F^{1/2}m_Z~\gamma_\mu\left[(g_{R}^{\ell} + \delta g_{R}^{\ell})P_R+(g_{L}^{\ell} + \delta g_{L}^{\ell})P_L\right]
\end{equation}
where the tree-level SM values are $g_{R}^{\ell} = 2 s_Z^2$ and $g_{L}^{\ell} = g_{R}^{\ell} - 1$ universally for $\ell = e,\mu,\tau$, with $s_Z$ being the sine of the weak mixing angle in the renormalisation scheme we utilise. That scheme uses the precisely measured parameters $G_F$ (from muon decay), $m_Z$ and $\alpha(m_Z)$ as input, so the definition of the weak mixing angle then becomes\footnote{Details for the different ingredients that enter the calculation of these corrections in this scheme follow \cite{Dawson:1994fa} closely.}
\begin{equation}
s_Z^2c_Z^2\equiv \frac{\pi\alpha(m_Z)}{\sqrt{2}G_F m^2_Z}.
\end{equation}
The quantities $\delta g_{R}^{\ell}$ and $\delta g_{L}^{\ell}$ are the one-loop corrections induced by $k$ interactions, and are discussed below. 

To simplify our calculation we treat the leptons as massless, in which case there are no corrections to $G_F$. We also assume that the new scalar $k$ is more massive than the $Z$-boson. There are two types of vertex corrections: the first one depends on the Yukawa couplings of the doubly charged scalars and is purely right-handed. It arises from the triangle diagrams and the lepton wave-function renormalisation. The second type of correction is common to all leptons  arising from the renormalisation of $\alpha$, $m_Z$, $s_Z$, the $Z$ wave function renormalisation, and $Z-\gamma$ mixing. Some details of the calculation, including analytic expressions with the vertex dependence on the mass of $k$ for $\delta g_{R}^{e,\mu,\tau}$ and $\delta g_{L}^{e,\mu,\tau}$ are presented in~\ref{sec:AppB} for completeness. 

With these ingredients we find for the doubly-charged isosinglet scalar that\footnote{The generalization to lepton flavour violating $Z$ boson decays is straightforward. The lepton-flavour violating right-handed $Z$-boson couplings are
$\delta g_{R}^{ij} = (y^\dagger y)_{ij}\, \delta r_{A}(m_{k}) + \Delta r(m_{k})$, 
where the first (second) index $i$ ($j$) denotes the (anti-)lepton flavour in the final state.
} 
\begin{align}\label{eq:verres}
    \delta g_{R}^{\ell} &= (y^\dagger y)_{\ell\ell}\, \delta r_{A}(m_{k}) + \Delta r(m_{k}).
\end{align}
The left-handed $Z$-boson couplings are 
\begin{equation}
\delta g_L^{\ell}=\Delta l(m_k)
\end{equation}
with the explicit formula for $\Delta l(m_k)$ given in Eq.~\eqref{eq:delta_l}. Numerically, only the terms with $\delta r_A(m_k)$ are important when their corresponding coupling constants $y$ are of order one.

\begin{figure}[tb!]\centering
\includegraphics[width=0.7\linewidth]{ 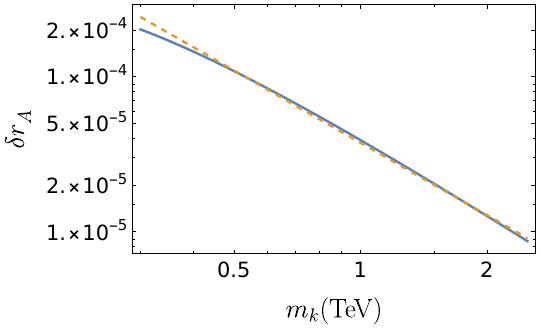}
\caption{$\delta r_A$ as a function of the scalar mass $m_k$. The dashed orange line shows the approximation in Eq.~\eqref{eq:dgr} to the full numerical result of $\delta r_A$ (blue solid line).}
\label{fig:dgr}
\end{figure}
For scalar masses $m_k\in [0.1,5]$ TeV, the function $\delta r_A$ ranges from $10^{-3}$ to $10^{-6}$ in magnitude. Current experiments are only sensitive to the real part of $\delta g_R^\ell$, which interferes with the SM contribution. The real part of $\delta g_R^\ell$ is well approximated by 
\begin{equation}\label{eq:dgr}
    \mathrm{Re}(\delta g_R^\ell) = (y^\dagger y)_{\ell \ell}\, \mathrm{Re}(\delta r_A), \qquad \mathrm{Re}(\delta r_A) = 3.75\times 10^{-5} \left(\frac{m_k}{1\,\mathrm{TeV}}\right)^{-1.55}\;,
\end{equation}
which is illustrated in Fig.~\ref{fig:dgr}.

The corrections from renormalisation that give rise to $\Delta r(m_{k})$ and $\Delta l(m_{k})$ affect many other observables as well and as such they should be constrained by a global fit. However, they are numerically insignificant, having values at most of the order of $10^{-6}$ for $m_{k}$ in the regime of interest. 
A comparison with the values extracted from LEP plus SLD data assuming lepton universality~\cite{ALEPH:2005ab}, 
\begin{equation}
g_{A}^{\ell}=-0.50123 \pm 0.00026,\qquad g_{V}^{\ell}=-0.03783 \pm 0.00041,
\label{eq:FUfit}
\end{equation}
confirms that these terms are indeed negligible for values of $m_{k}$ above 300 GeV.\footnote{Note that $g_A=(g_L-g_R)/2$ and $g_V=(g_L+g_R)/2$.} Since the SM obeys lepton flavour universality, in the following analysis of the flavour universality violating effects we take the comparison best fit value to be given by Eq.~\eqref{eq:FUfit}.

\begin{figure}[tb!]
\centering
\includegraphics[width=0.8\linewidth]{ 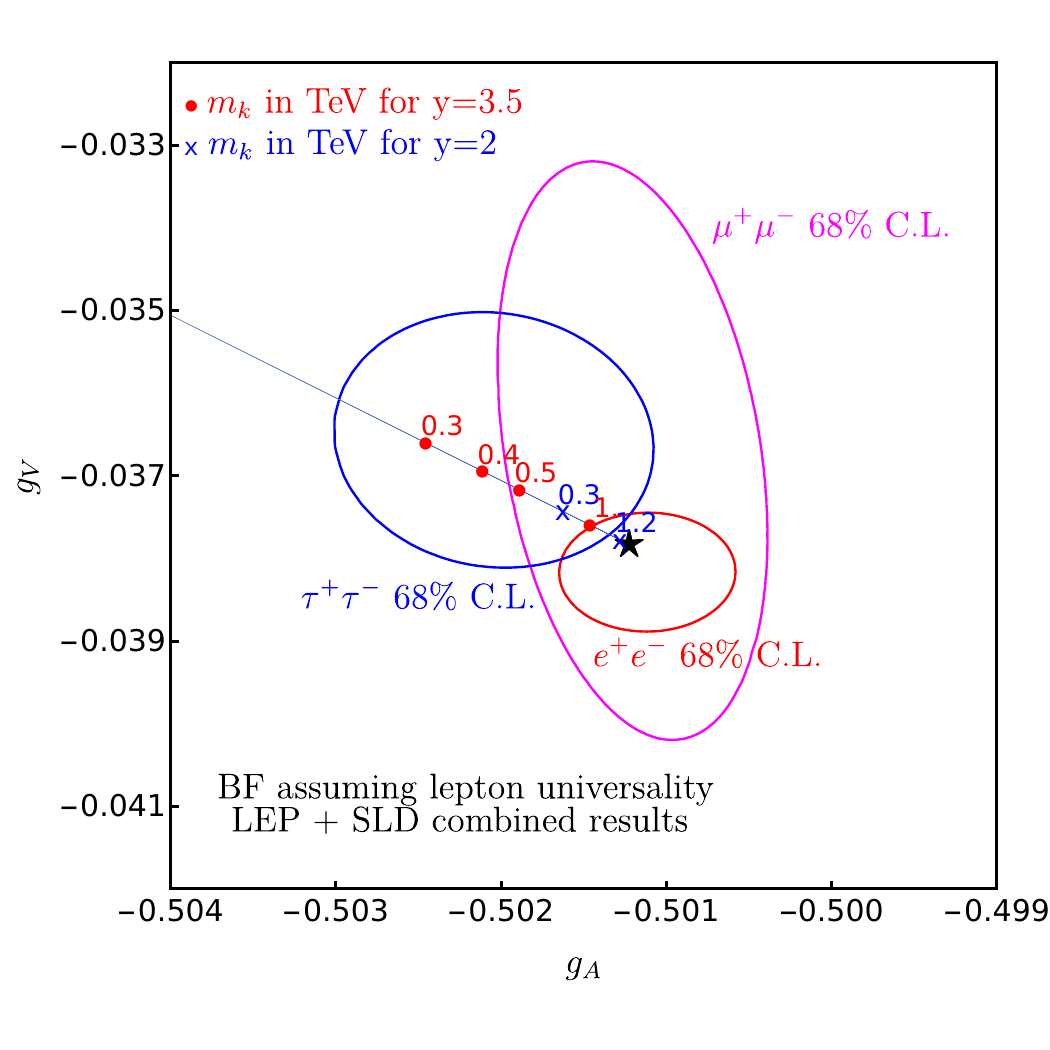}
\caption{LEP plus SLD allowed regions shown as the ellipse contours at 68$\%$ C.L.. The black star is the lepton universal best fit (BF) point~\cite{ALEPH:2005ab}.
Lepton flavour universality violating corrections to $Z\ell\ell$ shown by the thin blue line.
The real parts of $g_A$ and $g_V$ are plotted. Blue crosses (red circles) along this line mark illustrative values of $m_{k}$ in TeV for $y=2$ ($y=3.5$). The coupling constant $y$ in this case generically refers to 
$y\equiv\sqrt{(y^\dagger y)_{\ell\ell}}$ which appears in Eq.~\eqref{eq:verres}. 
}

\label{f:LEPNU}
\end{figure}

To constrain the flavour universality violating terms proportional to one of the Yukawa coupling combinations $(y^\dagger y)_{\ell\ell}$, we add them to the SM contribution and compare with the values given by the combined LEP and SLD data without the assumption of lepton universality, namely~\cite{ALEPH:2005ab}
\begin{equation}
\begin{aligned}
 g_{L}^{e}&= -0.26963\pm 0.00030, &  g_{R}^{e} &= 0.23148\pm 0.00029 ,\\
 g_{L}^{\mu}&= -0.2689\pm 0.0011, & g_{R}^{\mu}&= 0.2323\pm 0.0013,\\
g_{L}^{\tau}&= -0.26930 \pm 0.00058, &  g_{R}^{\tau}&= 0.23274 \pm 0.00062.
\label{eq:NFUfitLR}
\end{aligned}
\end{equation}

The results are illustrated in Fig.~\ref{f:LEPNU}, where we include only the real parts of the effective couplings. The imaginary parts do not interfere with the leading order SM contribution and thus enter at a higher order. We have checked numerically that their effect is small. The oval contours show the 68\% C.L.\ allowed regions corresponding to the combined LEP and SLD fit, with the lepton universal best fit point marked by the black star~\cite{ALEPH:2005ab}.\footnote{Note that the oval contours account for correlations, unlike the ranges quoted in Eq.~\eqref{eq:NFUfitLR}.} The thin blue line extending from the star illustrates the corrections arising from the universality violating contributions. The physical meaning of this line depends on which coupling is considered. 
We have marked several parameter points on this line to illustrate the sensitivity for different couplings $y$ and doubly-charged scalar masses $m_k$. The figure shows that switching on the doubly-charged scalar Yukawa couplings tends to worsen the fit to the $e^+ e^-$ data but improve the fit to the $\tau^+ \tau^-$ data. We emphasise that while there is no significant tension between the data in the three channels and the SM, the doubly-charged scalar Yukawa couplings produce an interesting flavour-dependent effect with a systematic trend.

\begin{figure}[tp!]
    \centering
\begin{tikzpicture}

\node[inner sep=0pt]  at (-2.35,11)
    {\includegraphics{ 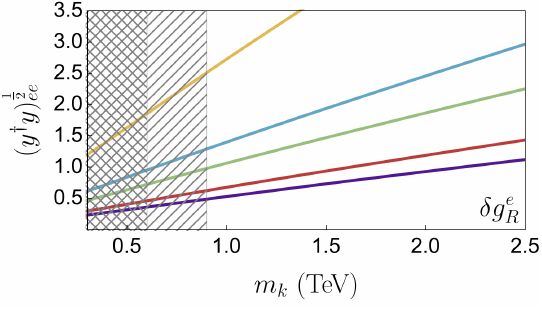}};
\node[inner sep=0pt] at (0,5)
    {\includegraphics[trim={0 0 .1cm 0},clip]{ 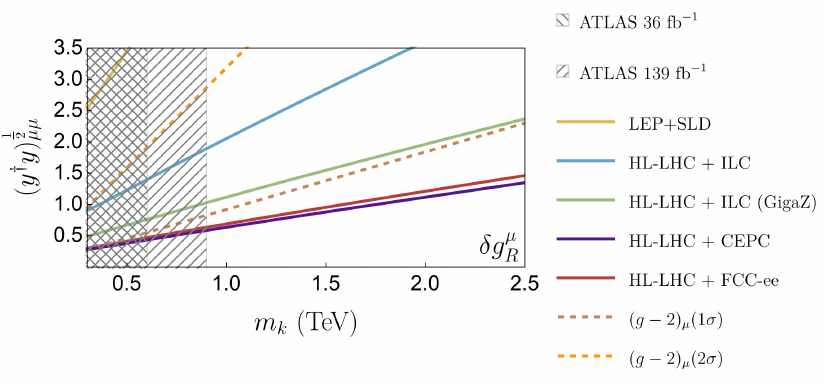}};
    \node[inner sep=0pt]  at (-2.35,-1)
    {\includegraphics{ 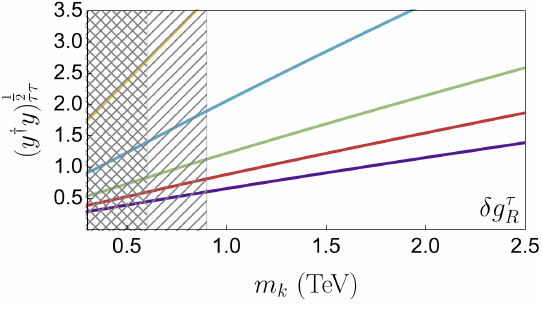}};

\end{tikzpicture}
    \caption{Constraints and parameter-space reach for the leptonic $Z$ boson couplings $\delta g_R^\ell$, with $\ell = e, \mu, \tau$. According to Eqs.~\eqref{eq:verres1A} to \eqref{eq:verres3A}, for the $T=(1,2,3)$ model, $\delta g_R^e$ varies with ($f_2, g_1, h_1$) respectively, $\delta g_R^\mu$ varies with ($f_1, g_2, h_1$), and $\delta g_R^\tau$ varies with ($f_1, g_1, h_2$). The mass of the doubly charged scalar $m_{k}$ is constrained by direct searches at the LHC to be larger than 600 GeV (ATLAS 36 fb$^{-1}$), or 900 GeV for universal lepton couplings (ATLAS 139 fb$^{-1}$), as discussed in Sec.~\ref{sec:existing-constraints} and shown by the grey hatched regions.  The region below the yellow contour corresponds to the allowed parameter space from the existing LEP and SLD bounds, with the uncertainty  at the 61\% level~\cite{ALEPH:2005ab}.
    The large improvement in sensitivity at the future colliders HL-LHC combined with ILC, ILC (Giga-Z), CEPC, and FCC-ee is illustrated by the blue, green, purple, and red lines, respectively; the uncertainties are at the $1\sigma$ level.  
}
    \label{fig:future}
\end{figure}

Future colliders will provide improved sensitivity to the leptonic $Z$-boson couplings. We now consider the sensitivity of four future colliders, with parameters extracted from Ref.~\cite{deBlas:2019rxi}: the HL-LHC~\cite{Cepeda:2019klc} with integrated luminosity of 6  ab$^{-1}$ in 12 years; the ILC 
at 250 GeV with 2 ab$^{-1}$; CEPC 
over 10 years for three different centre-of-mass energies: $\sqrt{s} = m_Z$  (60 ab$^{-1}$), $\sqrt{s}=2 m_W$ (3.6  ab$^{-1}$), and 240 GeV (12 ab$^{-1}$);  and the Future Circular Collider 
in the electron-positron configuration (FCC-ee) with runs at centre-of-mass energies $m_Z$ (150 ab$^{-1}$), $2m_W$ (10 ab$^{-1}$), 240 GeV (5 ab$^{-1}$) and 365 GeV (1.5 ab$^{-1}$) over a period of 13 years. This will amount to $5\times 10^{12}$ $Z$ bosons, an improvement of more than five orders of magnitude over LEP which recorded $1.7\times 10^7$ $Z$ bosons~\cite{ALEPH:2005ab}.\footnote{SLD produced $6\times 10^5$ $Z$ bosons.} Moreover, a $\sqrt{s}=m_Z$ $Z$-pole run called ``Giga Z'' is possible at the ILC which would drastically improve the experimental sensitivity. We analyse this for an integrated luminosity of 100 fb$^{-1}$ over a period of 1 to 3 years.

Figure~\ref{fig:future} shows the relevant parameter space region for the Yukawa coupling combination $(y^\dagger y)^{1/2}_{\ell\ell}$ 
as a function of the doubly-charged scalar mass $m_{k}$. It
uses the experimental uncertainties on the leptonic $Z$ boson couplings $\delta g_R^{e, \mu, \tau}$ given in Table~\ref{tab:gRl} to either constrain doubly-charged scalar boson parameter space (existing LEP+SLD data) or show the expected sensitivity of the combination of HL-LHC and the other future colliders we are considering. These results are derived from Eq.~\eqref{eq:verres}
noting $\Delta r(m_{k})$ is negligible. The experimental uncertainties for the current best measurements from LEP+SLD have been 
taken from Table 7.7 in~\cite{ALEPH:2005ab} (as also quoted in Eq.~\eqref{eq:NFUfitLR}) and the sensitivities of the future lepton colliders combined with the HL-LHC  are extracted from 
Figure 8 in Ref.~\cite{Belloni:2022due}.\footnote{Note the different definitions of the parameters $\delta g$. While we use the absolute uncertainty, Ref.~\cite{Belloni:2022due} uses a relative uncertainty.} 
The yellow contour is the upper edge of the currently allowed parameter space from LEP+SLD data, while the blue, green, purple and red lines are projections for the lower edge of the parameter reach of the HL-LHC combined with ILC, ILC (Giga-Z), CEPC and FCC-ee respectively. The constraints are approximately proportional to $m_k^{0.775}$, which orginates from the functional form of $\delta r_A$, see Eq.~\eqref{eq:dgr}. For comparison we also include the constraint from the muon anomalous magnetic moment measurement as dashed lines. It is stronger than the current LEP+SLD bound and even outperforms the combination of HL-LHC with the ILC (GigaZ). The electron anomalous magnetic moment measurements are currently not competitive. 
The approximate exclusion limits from direct searches at ATLAS are indicated by the hatched regions, with labels of ``ATLAS 36 fb$^{-1}$''  and ``ATLAS 139 fb$^{-1}$'' as per the discussion in Sec.~\ref{sec:triality}.

\begin{table}[t]
    \centering
   \begin{tabular}{
      p{0.1\linewidth}
      >{\centering}p{0.14\linewidth}
      >{\centering}p{0.14\linewidth}
      >{\centering}p{0.14\linewidth}
      >{\centering}p{0.14\linewidth}
      >{\centering\arraybackslash}p{0.14\linewidth}
      } \toprule
    \renewcommand{\arraystretch}{2}
      &LEP+SLD&ILC&ILC (Giga Z)&CEPC&FCC-ee\\
      &&+HL-LHC&+HL-LHC&+HL-LHC&+HL-LHC \\
      \midrule

  $\delta g_R^e $   $[\times 10^{4}]$& 2.9 & 0.76 & 0.44 & 0.11 & 0.18 \\
      $\delta g_R^\mu $   $[\times 10^{4}]$& 13.0 & 1.64 & 0.48 & 0.16 & 0.18 \\

       $\delta g_R^\tau $   $[\times 10^{4}]$& 6.2 & 1.65 & 0.58 & 0.17 & 0.30 \\
  
         \bottomrule
    \end{tabular}
 
    \caption{
    Existing or expected experimental precision for the leptonic $Z$-boson gauge couplings (multiplied by $10^4$). The existing experimental uncertainty for LEP+SLD was taken from Table 7.7 in Ref.~\cite{ALEPH:2005ab} 
    while the sensitivities for the combination of the future lepton colliders with the HL-LHC have been 
    extracted from Figure 8 in Ref.~\cite{Belloni:2022due}. 
    Giga Z refers to a $Z$ pole run, i.e.~$\sqrt{s}=m_Z$, at the ILC. The uncertainty of LEP+SLD is at the 61\% level~\cite{ALEPH:2005ab} and the future senstitivities are at the $1\sigma$ level.  
    Note that despite the large increase in statistics, the improved reach expected at the future colliders is limited by systematic uncertainties -- see e.g.~Ref.~\cite{deBlas:2019rxi}.
    }\label{tab:gRl}
\end{table}

The results straightforwardly translate to the lepton triality models. 
The corrections to the right-handed $Z$-boson couplings can be explicitly written as 
\begin{align}\label{eq:verres1A}
    \delta g_{R}^{\mu, \tau} &=f_1^2\, \delta r_{A}(m_{k_1}) + \Delta r(m_{k_1}), &
    \delta g_{R}^{e} &=f_2^2\, \delta r_{A}(m_{k_1})+ \Delta r(m_{k_1}),
\end{align}
in the $T=1$ model, 
whereas for the $T=2$ model we have that
\begin{align}\label{eq:verres2A}
    \delta g_{R}^{e,\tau} &=g_1^2\, \delta r_{A}(m_{k_2}) + \Delta r(m_{k_2}), &
    \delta g_{R}^{\mu} &=g_2^2\, \delta r_{A}(m_{k_2}) + \Delta r(m_{k_2}),
\end{align}
and for the $T=3$ model,
\begin{align}
    \delta g_R^{e,\mu} &=h_1^2\, \delta r_{A}(m_{k_3}) + \Delta r(m_{k_3}), &
    \delta g_{R}^{\tau} &=h_2^2\, \delta r_{A}(m_{k_3}) + \Delta r(m_{k_3}).
    \label{eq:verres3A}
\end{align}
Hence $\delta g_R^e$ varies with $(f_2,g_1,h_1)$, $\delta g_R^\mu$ varies with $(f_1,g_2,h_1)$ and $\delta g_R^\tau$ varies with $(f_1,g_1,h_2)$.

\subsection{\texorpdfstring{$H \to \gamma \gamma$}{H->gamma gamma} and \texorpdfstring{$H \to Z\gamma$}{H->Z gamma}}

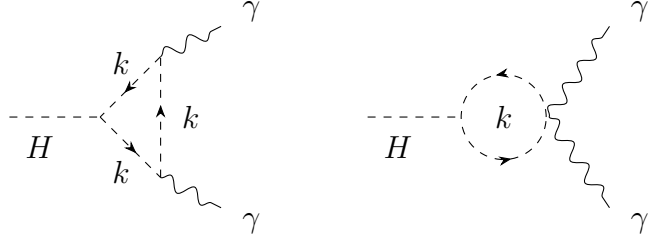
\begin{figure}[tb!]
    \centering
     \begin{tikzpicture}[scale =0.4,
  ->-/.style={decoration={markings,mark=at position 0.6 with {\arrow{Stealth}}},postaction={decorate}},]
  \filldraw[black] (0,-1) node{$H$};
 \draw [-,black, dashed] (-1,0) -- (2,0);
 
 \draw[black, dashed,->-] (2,0) -- (4,-2);
 \filldraw[black] (2.7,-1.8) node{$k$};
 \draw[black,  decorate,decoration=snake] (4,-2) -- (6,-3);
 \filldraw[black] (7,-3.5) node{$\gamma$};
 
 \draw[black, dashed,->-](4,-2)--(4,2);
\filldraw[black] (5,0)  node{$k$};

 \draw[black, dashed,->-] (4,2) -- (2,0);
 \filldraw[black] (2.7,1.8) node{$k$};
\draw[black,  decorate,decoration=snake] (4,2) -- (6,3);
 \filldraw[black] (7,3.5) node{$\gamma$};
\end{tikzpicture}
\hspace{1cm}
\begin{tikzpicture}[scale =0.4,
  ->-/.style={decoration={markings,mark=at position 0.6 with {\arrow{Stealth}}},postaction={decorate}},
  ->>-/.style={decoration={markings,mark=between positions 0.28 and 0.78 step 0.5 with {\arrow{Stealth}}},postaction={decorate}},
  ]

\filldraw[black] (9,-1) node{$H$};
 \draw [-,black, dashed] (8,0) -- (11,0);

 \draw[black,  decorate,decoration=snake] (14,0) -- (16,-3);
 \filldraw[black] (17,-3.5) node{$\gamma$};
 
1
\draw[black,  decorate,decoration=snake] (14,0) -- (16,3);
 \filldraw[black] (17,3.5) node{$\gamma$};

\draw[dashed,->>-](12.5,0)circle(40pt);
 \filldraw[black] (12.5,0) node{$k$};

 \end{tikzpicture}
\caption{The doubly-charged scalar contributions to $H\to \gamma \gamma$.}
    \label{f:Hgg-diagrams}
\end{figure}

We now return to the general model. After electroweak symmetry breaking, the Higgs portal coupling produces a cubic interaction between the physical Higgs boson and the doubly charged scalar:
\begin{equation}
\kappa_\phi\, \phi^\dagger \phi\, |k|^2 \supset \kappa_\phi v H |k|^2
\end{equation}
where $\phi^0 = (v + H)/\sqrt{2}$ in unitary gauge. 
This interaction produces a BSM contribution to the two-photon decay of the Higgs boson from doubly-charged scalar loops, with the vertex corrections shown in Fig.~\ref{f:Hgg-diagrams}. Relative to the SM the $H\to\gamma\gamma$ rate becomes,
\begin{align}
R_{\gamma\gamma}=&\frac{\Gamma(H\to\gamma\gamma)}{ \Gamma(H\to\gamma\gamma)_{SM}}=\left|\frac{I_V\left(\tau_W\right)+\frac{4}{3}I_q\left(\tau_t\right)+2\kappa_\phi\frac{v^2}{m_{k_i}^2} I_S\left(\tau_{k}\right)}{I_V\left(\tau_W\right)+\frac{4}{3}I_q\left(\tau_t\right)}\right|^2
\end{align}
where $I_V$ and $I_q$ are the well-known loop factors for the $W$ and $t$-quark SM contributions, and $I_S$ is the corresponding loop function for the new contribution due to the $k$ scalar. The arguments are given by $\tau_a = 4m_a^2/m_H^2$ where $a = W, t, k$, with $m_H$ being the mass of the Higgs boson. The loop functions are collected in~\ref{sec:AppA} for reference. This result agrees with Ref.~\cite{Das:2016bir} when the latter is specialised to couplings pertaining to our model.

\begin{figure}[t!]
    \centering
\begin{tikzpicture}     \node (pic) {\includegraphics[width = 0.7\textwidth]{ 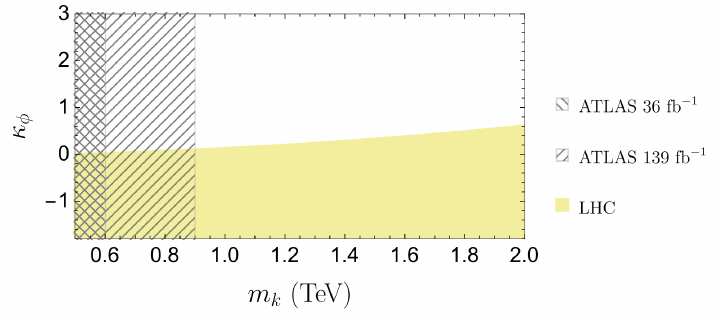}};
\node[below right] at (pic.south west){ \includegraphics[width = 0.8\textwidth]{ 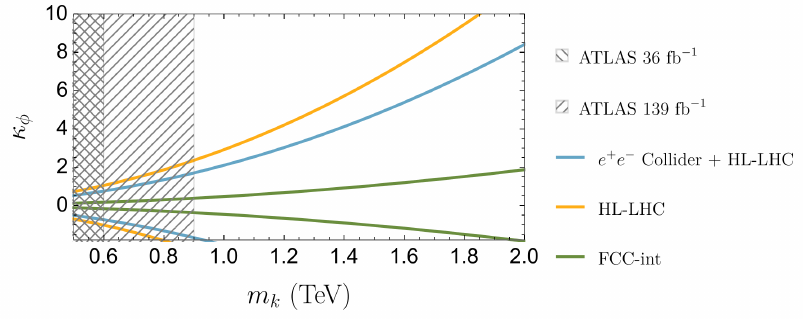}};
\end{tikzpicture}

    \caption{
    Experimental sensitivity to the Higgs portal coupling $\kappa_\phi$ from precise measurements of $H \to \gamma \gamma$ as a function of $m_{k}$. Both panels show the current ATLAS indicative constraints from direct searches as exclusion regions in grey hatch. The yellow region in the top panel shows the currently allowed region from the Higgs diphoton channel. The lower panel shows the projections for HL-LHC with $6\,\mathrm{ab}^{-1}$ of data in orange, a future $e^+ e^-$ collider running at 240 GeV taking 5  ab$^{-1}$ of data in blue combined with HL-LHC (see text for discussion), and the integrated FCC program ($ee$, $eh$ and $hh$) in green. The colliders are able to probe the regions outside the contours.   }
    \label{fig:fig6New}
\end{figure}

In the top panel of Fig.~\ref{fig:fig6New} the yellow-shaded area shows the region of $m_{k}-\kappa_\phi$ parameter space that is currently allowed at $1\sigma$ by the ATLAS Run 2 measurements, $R_{\gamma\gamma}=1.088^{+0.095}_{-0.09}$ \cite{ATLAS:2022vkf}, assuming no new physics in the production process. 
We show values of $\kappa_\phi$ in the range $[-1.8,10]$ which are consistent with perturbative unitarity and the bounded-from-below bounds discussed in Sec.~\ref{sec:doubly-charged}.
Note that the current measurements favour BSM contributions that increase the rate compared to the SM, whereas in general BSM can also reduce the rate. This occurs for negative values of $\kappa_\phi$ in our models.

In the lower panel of Fig.~\ref{fig:fig6New} we show the expected sensitivities of the future colliders using parameters extracted from Table 6 in Ref.~\cite{Dawson:2022zbb}. 
The region bounded by the orange contour illustrates the future sensitivity of HL-LHC, derived from the expected precision $\Delta\kappa_\gamma=1.8\%$  where $R_{\gamma\gamma}=\kappa_\gamma^2$.
The blue line illustrates the sensitivity of an $e^+e^-$ collider combined with HL-LHC, where we suppose a centre-of-mass energy of 240 GeV and 5  ab$^{-1}$ of data, which leads to $\Delta\kappa_\gamma = 1.3 \%$, corresponding to the FCC-ee sensitivity. However, similar results may be achieved by CEPC ($\Delta\kappa_\gamma = 1.68 \%$) and ILC  ($\Delta\kappa_\gamma = 1.36 \%$) with the same centre-of-mass energy of 240 GeV; both are also combined with HL-LHC. A single line thus suffices to illustrate all proposed $e^+ e^-$ colliders. 
 Finally, the green contour pertains to the full integrated FCC program ($ee$, $eh$ and $hh$) with $\Delta\kappa_\gamma=0.29\%$. Note that FCC-int is not combined with HL-LHC.
 A detailed discussion of the luminosities for each run of the FCC, CEPC and ILC programs are described in Sec.~\ref{sub:Zll}.

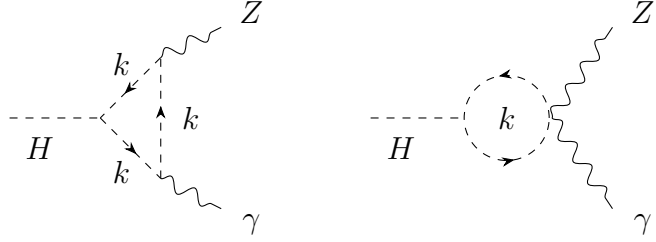
\begin{figure}[t]
    \centering
     \begin{tikzpicture}[scale =0.4,
  ->-/.style={decoration={markings,mark=at position 0.6 with {\arrow{Stealth}}},postaction={decorate}},]
  \filldraw[black] (0,-1) node{$H$};
 \draw [-,black, dashed] (-1,0) -- (2,0);
 
 \draw[black, dashed,->-] (2,0) -- (4,-2);
 \filldraw[black] (2.7,-1.8) node{$k$};
 \draw[black,  decorate,decoration=snake] (4,-2) -- (6,-3);
 \filldraw[black] (7,-3.5) node{$\gamma$};
 
 \draw[black, dashed,->-](4,-2)--(4,2);
\filldraw[black] (5,0)  node{$k$};

 \draw[black, dashed,->-] (4,2) -- (2,0);
 \filldraw[black] (2.7,1.8) node{$k$};
\draw[black,  decorate,decoration=snake] (4,2) -- (6,3);
 \filldraw[black] (7,3.5) node{$Z$};
\end{tikzpicture}
\hspace{1cm}
\begin{tikzpicture}[scale =0.4,
  ->-/.style={decoration={markings,mark=at position 0.6 with {\arrow{Stealth}}},postaction={decorate}},
  ->>-/.style={decoration={markings,mark=between positions 0.28 and 0.78 step 0.5 with {\arrow{Stealth}}},postaction={decorate}},
  ]

\filldraw[black] (9,-1) node{$H$};
 \draw [-,black, dashed] (8,0) -- (11,0);

 \draw[black,  decorate,decoration=snake] (14,0) -- (16,-3);
 \filldraw[black] (17,-3.5) node{$\gamma$};
 
\draw[black,  decorate,decoration=snake] (14,0) -- (16,3);
 \filldraw[black] (17,3.5) node{$Z$};

\draw[dashed,->>-](12.5,0)circle(40pt);
 \filldraw[black] (12.5,0) node{$k$};

 \end{tikzpicture}
 
 \caption{The doubly-charged scalar contributions to $H\to Z \gamma$. We do not show the self-energy corrections, but they are incorporated in the final result.}
     \label{f:HZg-diagrams}
\end{figure}

Following a very similar procedure to the above, we can write for the $H \to Z\gamma$ decay that
\begin{align}
R_{Z\gamma}=&\frac{\Gamma(H\to Z\gamma)}{ \Gamma(H\to Z\gamma)_{SM}}=\left|
\frac{A^{Z\gamma}_t+A^{Z\gamma}_W+A^{Z\gamma}_{k}}{A^{Z\gamma}_t+A^{Z\gamma}_W}
\right|^2
\end{align}
where
\begin{subequations}
\begin{align}
A_{t}^{Z\gamma}  &=\frac{-4(\frac{1}{2}-\frac{4}{3}\sin^2\theta_W)}{\sin\theta_W\cos\theta_W} \left[ I_1(\tau_t,\lambda_t)-I_2(\tau_t,\lambda_t)\right], \\
A_W^{Z\gamma} &=-\cot\theta_W \left[4(3-\tan^2\theta_W)I_2(\tau_W,\lambda_W)+[(1+\tfrac{2}{\tau_W})\tan^2 \theta_W-(5+\tfrac{2}{\tau_W})]I_1(\tau_W,\lambda_W)\right],\\
A_{k}^{Z\gamma} &=-4\kappa_\phi\tan\theta_W\frac{v^2}{m_{k}^2} I_1(\tau_{k},\lambda_{k}),
\end{align}
\end{subequations}
with $\lambda_a = 4m_{a}^2/m_Z^2$, and with the known one-loop functions $I_1$ and $I_2$ given in~\ref{sec:AppC} for reference. 
The SM top-quark and $W$-boson contributions ($A_{t}^{Z\gamma}$, $A_W^{Z\gamma}$) have been first calculated in Ref.~\cite{Bergstrom:1985hp} and are taken from Ref.~\cite{Gunion:1989we} and the scalar contribution is similar to that of the physical charged Higgs boson in two-Higgs-doublet models. The vertex corrections induced by $k$ interactions are shown in Fig.~\ref{f:HZg-diagrams}.

\begin{figure}[t]
\centering{
\includegraphics[width = 0.7\textwidth]{ 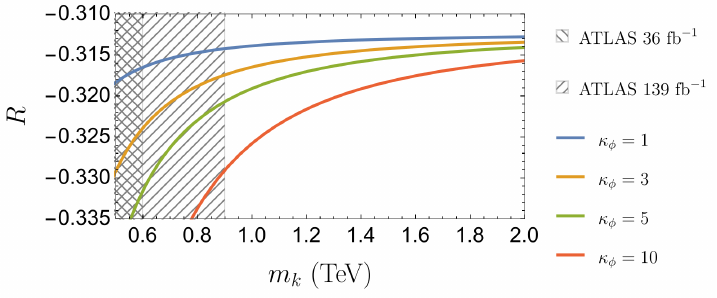}}
\caption{The ratio $R=(R_{Z\gamma}-1)/(R_{\gamma\gamma}-1)$ as a function of $m_{k_i}$ for selected values of $\kappa_\phi$. This demonstrates that $H \to Z \gamma$ is at best about an order of magnitude less sensitive to doubly-charged scalar loop effects compared to $H \to \gamma \gamma$.   }
\label{f:HZg}
\end{figure}

\begin{figure}[t]
    \centering
   \includegraphics[width = 0.7\textwidth]{ 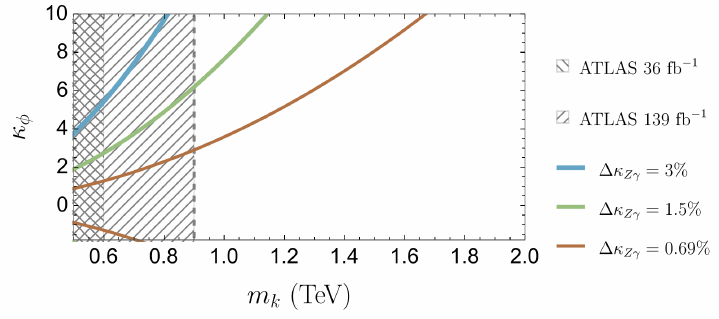}

    \caption{
    The anticipated HL-LHC precision of $\Delta \kappa_{Z\gamma} = 9.8\%$ for $R_{Z\gamma}$ will be insufficient to constrain the $(m_{k}, \kappa_\phi)$ parameter space in any meaningful way. The curves show the parameter space that could be probed by hypothetical precision measurements at the $\Delta \kappa_{Z\gamma} = 3\%$, $1.5\%$ and $0.69\%$. These regions lie outside the blue, green and brown curves, respectively. The integrated FCC ($ee$, $eh$ and $hh$) is expected to reach a $\Delta \kappa_{Z\gamma} = 0.69\%$ sensitivity~\cite{Bernardi:2022hny}.  }
    \label{fig:PrecisionZgamma}
\end{figure}

The ratio $R_{Z\gamma}=2.2\pm 0.7$ has been recently measured \cite{ATLAS:2022vkf,CMS:2023mku}\footnote{We do not attempt to explain the hint of an excess over the SM expectation -- see Refs.~\cite{Boto:2023bpg,Hong:2023mwr,He:2024bxi} for possible explanations.}, and the future precision at HL-LHC is expected to be 
$\Delta\kappa_{Z\gamma} = 9.8\%$ where $R_{Z\gamma}=\kappa_{Z\gamma}^2$ \cite{deBlas:2019rxi, Dainese:2019rgk,ATL-PHYS-PUB-2022-018, Mlynarikova:2023bvx}. Future lepton colliders cannot improve this measurement. The full FCC program with $ee/eh/hh$ running would, however, increase the precision to  $\Delta\kappa_{Z\gamma} = 0.69\%$~\cite{deBlas:2019rxi, Bernardi:2022hny}. Since this observable is currently measured less precisely than $R_{\gamma\gamma}$, and that situation is expected to continue into the future, it is not expected to provide competitive constraints on the parameter space. Nevertheless, it is interesting to quantify the effect of the BSM physics we are considering, which is done in Fig.~\ref{f:HZg}. We see that deviations from the SM would be at most at the $31 \,$-- $33\%$ level in $H \to Z \gamma$ compared to $H \to \gamma \gamma$. To understand the plot properly, note that in the limit $m_{k} \to \infty$, where the new physics decouples, this ratio goes to one-loop accuracy as:\footnote{Note that higher-order corrections are important in the SM. The ratio of branching ratios $\sqrt{\frac{{\cal B}(h\to \gamma\gamma)_{SM}}{{\cal B}(h\to Z\gamma)_{SM}}}=0.824$~\cite{LHCHiggsCrossSectionWorkingGroup:2016ypw} receives a $\sim 45\%$ correction over the leading-order (one-loop) prediction of $0.570$.}  
\begin{equation}
    \frac{R_{Z\gamma}-1}{R_{\gamma\gamma}-1} \approx - \tan\theta_W\left(1+\frac{1}{15}\frac{m_Z^2}{m_{k}^2} \right) 
    \sqrt{1-\frac{m_Z^2}{m_h^2}}
    \sqrt{\frac{{\cal B}(h\to \gamma\gamma)_{SM}}{{\cal B}(h\to Z\gamma)_{SM}}}.
\end{equation}
The anticipated $\Delta \kappa_{Z\gamma} = 9.8\%$ precision at HL-LHC will not be sufficient to constrain the $(m_{k_i}, \kappa_\phi)$ parameter space in any meaningful way. Figure~\ref{fig:PrecisionZgamma} shows the constraints that could be achieved with hypothetical precision measurements at the $\Delta \kappa_{Z\gamma} = 3\%$, $1.5\%$ and $0.69\%$ levels.

The results directly apply to the lepton triality models, because the Higgs portal coupling $\kappa$ is not affected by the lepton triality symmetry.
 
\section{Conclusion}
\label{sec:conc}
We revisited corrections to the leptonic $Z$-boson couplings and Higgs boson decays to two photons or a photon and a $Z$ boson in models with a doubly-charged scalar $k$. These general results have then been applied to to simple lepton triality models featuring doubly-charged scalar bosons $k_i$ which serve as useful benchmark BSM scenarios for charged lepton flavour violation. 

Precision measurements of the leptonic $Z$-boson couplings are sensitive to both flavour-preserving and flavour-changing Yukawa couplings of the doubly-charged scalar. We have found that the processes $Z \to \ell^+ \ell^-$ $(\ell = e, \mu, \tau)$, which conserve lepton flavour, exhibit interesting flavour non-universality effects due to one-loop level interactions of the doubly-charged scalar, as summarised in Figs.~\ref{f:LEPNU} and \ref{fig:future}. 
Figure~\ref{f:LEPNU} shows that doubly-charged scalars can improve the fit to the $Z\tau\tau$ coupling but at the expense of the $Zee$ coupling. 
Figure~\ref{fig:future} shows both existing bounds from LEP+SLD and the future parameter-space reach of HL-LHC when combined, respectively, with the ILC, ILC (Giga Z), CEPC, and FCC-ee which can be directly applied to the lepton triality specific cases. 

In addition, the Higgs portal coupling $\kappa_\phi$ between the doubly-charged scalar $k$ and the standard Higgs boson is constrained, as a function of $m_{k}$, by $H \to \gamma \gamma$, as depicted in Fig.~\ref{fig:fig6New}.  This figure also shows the future sensitivity at the HL-LHC and future $e^+e^- $ colliders. Effects on $H \to Z \gamma$ were also analysed, but this decay mode is not expected to provide useful constraints for the foreseeable future.

\section*{Acknowledgments}
We thank Fang Xu for communication about the anomalous magnetic moment. This work was supported in part by Australian Research Council Discovery Project DP200101470. 

\appendix
\mathversion{bold}
\section{\texorpdfstring{$Z\to \ell^+\ell^-$}{Z->l+l-}}
\mathversion{normal}
\label{sec:AppB}

To obtain these results we have implemented the model in  \texttt{FeynRules}~\cite{Christensen:2009jx,Alloul:2013bka} to generate \texttt{FeynArts}~\cite{Hahn:2000kx} output where \texttt{LoopTools}~\cite{Hahn:1998yk} is used to compute the one-loop diagrams and simplification is assisted with \texttt{FeynCalc}~\cite{Mertig:1990an,Shtabovenko:2016sxi,Shtabovenko:2020gxv,Shtabovenko:2023idz}, \texttt{FeynHelpers}~\cite{Shtabovenko:2016whf} and \texttt{Package-X}~\cite{Patel:2015tea,Patel:2016fam}.

We combine the different ingredients of the calculation to obtain,
\begin{align}
\delta r_{A}(m_{k}) = & \frac{s_Z^2}{32\pi^2 m_Z^2}\Bigl(-4m_{k}^4(C_0(0,0,m_Z^2,0,m_{k}^2,0)+2C_0(0,0,m_Z^2,m_{k}^2,0,m_{k}^2)) \nonumber\\
+& 4\frac{2m_{k}^2-m_Z^2}{m_Z} {\cal H}(m_{k},m_Z) -7m_Z^2+2(2m_{k}^2-m_Z^2)(1-i\pi-2\log(m_{k}/m_Z)) \Bigr),\label{eq:deltarA}\\
\Delta r(m_{k}) = & \frac{e^2s_Z^2}{36\pi^2c_Z^2(c_Z^2-s_Z^2)}\frac{1}{m_Z^3}\Bigl(
6({\cal H}(m_{k},m_Z)+2m_Z)((3s_Z^4-3s_Z^2c_Z^2+4c_Z^4)m_{k}^2-c_Z^4m_Z^2) \nonumber\\
-& m_Z^3(3s_Z^4-3s_Z^2c_Z^2+4c_Z^4) \Bigr), \label{eq:delta_r}\\
\Delta l(m_{k}) =& \frac{e^2s_Z^2}{36\pi^2c_Z^2(c_Z^2-s_Z^2)}\frac{1}{m_Z^3}\Bigl(
6({\cal H}(m_{k},m_Z)+2m_Z)((3s_Z^4-3s_Z^2c_Z^2+4c_Z^4)m_{k}^2-c_Z^4m_Z^2) \nonumber\\
-&m_Z^3(3s_Z^4-3s_Z^2c_Z^2+4c_Z^4) +(c_Z^2-s_Z^2)(9m_{k}^2({\cal H}(m_{k},m_Z)+2m_Z)-\frac{3}{2}m_Z^3)\Bigr).\label{eq:delta_l}
\end{align}
We have left our result in terms of the Passarino-Veltman function $C_0$ for compactness. The complete vertex corrections are constructed from these functions as in Eq.~\eqref{eq:verres}.
The loop function ${\cal H}$, can be written in terms of 
\begin{equation}
r^2=\frac{m_Z^2}{4m_{k}^2}
\end{equation}
for $r\leq 1$ as
\begin{equation}
{\cal H}(m_{k},m_Z)=-m_Z\frac{1}{r}\sqrt{1-r^2}\arccos(1-2r^2).
\end{equation}

\mathversion{bold}
\section{\texorpdfstring{$H\to \gamma \gamma$}{H->gamma gamma}}
\mathversion{normal}
\label{sec:AppA}

For convenience we list here the standard one-loop form factors that appear in $H\to\gamma\gamma$ for intermediate vectors, fermions and scalars, respectively:
\begin{align}
I_V(x) &= 2+3x+3x(2-x){\cal F}(x),\nonumber\\
I_F(x) &= -2x(1+(1-x ){\cal F}(x)),\nonumber\\ 
I_S(x) &= x(1-x{\cal F}(x)), 
\end{align}
where
\begin{align}
{\cal F}(x)=&\left\lbrace
\begin{array}{ccc}
-\dfrac{1}{4}\left(\ln\left(\dfrac{1+\sqrt{1-x}}{1-\sqrt{1-x}}\right)-i\pi\right)^2 & & \text{for }x<1\\ & & \\
\left(\arcsin\sqrt{\dfrac{1}{x}}\right)^2 & & \text{for }x\geq1
\end{array}\right. .
\end{align}

\mathversion{bold}
\section{\texorpdfstring{$H\to Z \gamma$}{H->Z gamma}}
\mathversion{normal}
\label{sec:AppC}

The one-loop form factors in this case are
\begin{align}
I_1(x,y) &= \frac{x y}{2(x-y)}+\frac{x^2 y^2}{2(x-y)^2}[{\cal F}(x)-{\cal F}(y)]+\frac{x^2 y}{(x-y)^2}[{\cal G}(x)-{\cal G}(y)],\\
I_2(x,y) &= -\frac{x y}{2(x-y)}[{\cal F}(x)-{\cal F}(y)],\,
\end{align}
where
\begin{eqnarray}
{\cal G}(x)=\left\lbrace
\begin{array}{ccc}
\dfrac{1}{2}\sqrt{1-x}\left(\ln\left(\dfrac{1+\sqrt{1-x}}{1-\sqrt{1-x}}\right)-i\pi\right) & & \text{for }x<1\\ & & \\
\sqrt{x-1}\arcsin\sqrt{\dfrac{1}{x}} & & \text{for }x\geq1
\end{array}\right. .
\end{eqnarray}

\bibliographystyle{elsarticle-harv}
\bibliography{refs}

\begin{thebibliography}{75}
\expandafter\ifx\csname natexlab\endcsname\relax\def\natexlab#1{#1}\fi
\providecommand{\url}[1]{\texttt{#1}}
\providecommand{\href}[2]{#2}
\providecommand{\path}[1]{#1}
\providecommand{\DOIprefix}{doi:}
\providecommand{\ArXivprefix}{arXiv:}
\providecommand{\URLprefix}{URL: }
\providecommand{\Pubmedprefix}{pmid:}
\providecommand{\doi}[1]{\href{http://dx.doi.org/#1}{\path{#1}}}
\providecommand{\Pubmed}[1]{\href{pmid:#1}{\path{#1}}}
\providecommand{\bibinfo}[2]{#2}
\ifx\xfnm\relax \def\xfnm[#1]{\unskip,\space#1}\fi
\bibitem[{Aaboud et~al.(2018)}]{ATLAS:2017xqs}
\bibinfo{author}{Aaboud, M.}, et~al. (\bibinfo{collaboration}{ATLAS}),
  \bibinfo{year}{2018}.
\newblock \bibinfo{title}{{Search for doubly charged Higgs boson production in
  multi-lepton final states with the ATLAS detector using
  proton\textendash{}proton collisions at $\sqrt{s}=13\,\text {TeV}$}}.
\newblock \bibinfo{journal}{Eur. Phys. J. C} \bibinfo{volume}{78},
  \bibinfo{pages}{199}.
\newblock \DOIprefix\doi{10.1140/epjc/s10052-018-5661-z},
  \href{http://arxiv.org/abs/1710.09748}{{\tt arXiv:1710.09748}}.
\bibitem[{Aad et~al.(2023a)}]{CMS:2023mku}
\bibinfo{author}{Aad, G.}, et~al. (\bibinfo{collaboration}{CMS, ATLAS}),
  \bibinfo{year}{2023}a.
\newblock \bibinfo{title}{{Evidence for the Higgs boson decay to a $Z$ boson
  and a photon at the LHC}}.
\newblock \href{http://arxiv.org/abs/2309.03501}{{\tt arXiv:2309.03501}}.
\bibitem[{Aad et~al.(2023b)}]{ATLAS:2022pbd}
\bibinfo{author}{Aad, G.}, et~al. (\bibinfo{collaboration}{ATLAS}),
  \bibinfo{year}{2023}b.
\newblock \bibinfo{title}{{Search for doubly charged Higgs boson production in
  multi-lepton final states using 139~fb$^{-1}$ of proton\textendash{}proton
  collisions at $\sqrt{s}$ = 13~TeV with the ATLAS detector}}.
\newblock \bibinfo{journal}{Eur. Phys. J. C} \bibinfo{volume}{83},
  \bibinfo{pages}{605}.
\newblock \DOIprefix\doi{10.1140/epjc/s10052-023-11578-9},
  \href{http://arxiv.org/abs/2211.07505}{{\tt arXiv:2211.07505}}.
\bibitem[{Abada et~al.(2019)}]{FCC:2018byv}
\bibinfo{author}{Abada, A.}, et~al. (\bibinfo{collaboration}{FCC}),
  \bibinfo{year}{2019}.
\newblock \bibinfo{title}{{FCC Physics Opportunities}: {Future Circular
  Collider Conceptual Design Report Volume 1}}.
\newblock \bibinfo{journal}{Eur. Phys. J. C} \bibinfo{volume}{79},
  \bibinfo{pages}{474}.
\newblock \DOIprefix\doi{10.1140/epjc/s10052-019-6904-3}.
\bibitem[{de~Adelhart~Toorop et~al.(2011a)de~Adelhart~Toorop, Bazzocchi, Merlo
  and Paris}]{deAdelhartToorop:2010jxh}
\bibinfo{author}{de~Adelhart~Toorop, R.}, \bibinfo{author}{Bazzocchi, F.},
  \bibinfo{author}{Merlo, L.}, \bibinfo{author}{Paris, A.},
  \bibinfo{year}{2011}a.
\newblock \bibinfo{title}{{Constraining Flavour Symmetries At The EW Scale I:
  The A4 Higgs Potential}}.
\newblock \bibinfo{journal}{JHEP} \bibinfo{volume}{03}, \bibinfo{pages}{035}.
\newblock \DOIprefix\doi{10.1007/JHEP03(2011)035},
  \href{http://arxiv.org/abs/1012.1791}{{\tt arXiv:1012.1791}}.
  \bibinfo{note}{[Erratum: JHEP 01, 098 (2013)]}.
\bibitem[{de~Adelhart~Toorop et~al.(2011b)de~Adelhart~Toorop, Bazzocchi, Merlo
  and Paris}]{deAdelhartToorop:2010nki}
\bibinfo{author}{de~Adelhart~Toorop, R.}, \bibinfo{author}{Bazzocchi, F.},
  \bibinfo{author}{Merlo, L.}, \bibinfo{author}{Paris, A.},
  \bibinfo{year}{2011}b.
\newblock \bibinfo{title}{{Constraining Flavour Symmetries At The EW Scale II:
  The Fermion Processes}}.
\newblock \bibinfo{journal}{JHEP} \bibinfo{volume}{03}, \bibinfo{pages}{040}.
\newblock \DOIprefix\doi{10.1007/JHEP03(2011)040},
  \href{http://arxiv.org/abs/1012.2091}{{\tt arXiv:1012.2091}}.
\bibitem[{Aguillard et~al.(2023)}]{Muong-2:2023cdq}
\bibinfo{author}{Aguillard, D.P.}, et~al. (\bibinfo{collaboration}{Muon g-2}),
  \bibinfo{year}{2023}.
\newblock \bibinfo{title}{{Measurement of the Positive Muon Anomalous Magnetic
  Moment to 0.20~ppm}}.
\newblock \bibinfo{journal}{Phys. Rev. Lett.} \bibinfo{volume}{131},
  \bibinfo{pages}{161802}.
\newblock \DOIprefix\doi{10.1103/PhysRevLett.131.161802},
  \href{http://arxiv.org/abs/2308.06230}{{\tt arXiv:2308.06230}}.
\bibitem[{Alloul et~al.(2014)Alloul, Christensen, Degrande, Duhr and
  Fuks}]{Alloul:2013bka}
\bibinfo{author}{Alloul, A.}, \bibinfo{author}{Christensen, N.D.},
  \bibinfo{author}{Degrande, C.}, \bibinfo{author}{Duhr, C.},
  \bibinfo{author}{Fuks, B.}, \bibinfo{year}{2014}.
\newblock \bibinfo{title}{{FeynRules 2.0 - A complete toolbox for tree-level
  phenomenology}}.
\newblock \bibinfo{journal}{Comput. Phys. Commun.} \bibinfo{volume}{185},
  \bibinfo{pages}{2250--2300}.
\newblock \DOIprefix\doi{10.1016/j.cpc.2014.04.012},
  \href{http://arxiv.org/abs/1310.1921}{{\tt arXiv:1310.1921}}.
\bibitem[{Altarelli and Feruglio(2006)}]{Altarelli:2005yx}
\bibinfo{author}{Altarelli, G.}, \bibinfo{author}{Feruglio, F.},
  \bibinfo{year}{2006}.
\newblock \bibinfo{title}{{Tri-bimaximal neutrino mixing, A(4) and the modular
  symmetry}}.
\newblock \bibinfo{journal}{Nucl. Phys. B} \bibinfo{volume}{741},
  \bibinfo{pages}{215--235}.
\newblock \DOIprefix\doi{10.1016/j.nuclphysb.2006.02.015},
  \href{http://arxiv.org/abs/hep-ph/0512103}{{\tt arXiv:hep-ph/0512103}}.
\bibitem[{Anisha et~al.(2022)Anisha, Banerjee, Chakrabortty, Englert,
  Spannowsky and Stylianou}]{Anisha:2021jlz}
\bibinfo{author}{Anisha}, \bibinfo{author}{Banerjee, U.},
  \bibinfo{author}{Chakrabortty, J.}, \bibinfo{author}{Englert, C.},
  \bibinfo{author}{Spannowsky, M.}, \bibinfo{author}{Stylianou, P.},
  \bibinfo{year}{2022}.
\newblock \bibinfo{title}{{Effective connections of a\ensuremath{\mu}, Higgs
  physics, and the collider frontier}}.
\newblock \bibinfo{journal}{Phys. Rev. D} \bibinfo{volume}{105},
  \bibinfo{pages}{016019}.
\newblock \DOIprefix\doi{10.1103/PhysRevD.105.016019},
  \href{http://arxiv.org/abs/2108.07683}{{\tt arXiv:2108.07683}}.
\bibitem[{Anisha et~al.(2023)Anisha, Das~Bakshi, Banerjee, Biek\"otter,
  Chakrabortty, Kumar~Patra and Spannowsky}]{Anisha:2021hgc}
\bibinfo{author}{Anisha}, \bibinfo{author}{Das~Bakshi, S.},
  \bibinfo{author}{Banerjee, S.}, \bibinfo{author}{Biek\"otter, A.},
  \bibinfo{author}{Chakrabortty, J.}, \bibinfo{author}{Kumar~Patra, S.},
  \bibinfo{author}{Spannowsky, M.}, \bibinfo{year}{2023}.
\newblock \bibinfo{title}{{Effective limits on single scalar extensions in the
  light of recent LHC data}}.
\newblock \bibinfo{journal}{Phys. Rev. D} \bibinfo{volume}{107},
  \bibinfo{pages}{055028}.
\newblock \DOIprefix\doi{10.1103/PhysRevD.107.055028},
  \href{http://arxiv.org/abs/2111.05876}{{\tt arXiv:2111.05876}}.
\bibitem[{Aoyama et~al.(2020)}]{Aoyama:2020ynm}
\bibinfo{author}{Aoyama, T.}, et~al., \bibinfo{year}{2020}.
\newblock \bibinfo{title}{{The anomalous magnetic moment of the muon in the
  Standard Model}}.
\newblock \bibinfo{journal}{Phys. Rept.} \bibinfo{volume}{887},
  \bibinfo{pages}{1--166}.
\newblock \DOIprefix\doi{10.1016/j.physrep.2020.07.006},
  \href{http://arxiv.org/abs/2006.04822}{{\tt arXiv:2006.04822}}.
\bibitem[{Aryshev et~al.(2022)}]{ILCInternationalDevelopmentTeam:2022izu}
\bibinfo{author}{Aryshev, A.}, et~al. (\bibinfo{collaboration}{ILC
  International Development Team}), \bibinfo{year}{2022}.
\newblock \bibinfo{title}{{The International Linear Collider: Report to
  Snowmass 2021}} \href{http://arxiv.org/abs/2203.07622}{{\tt
  arXiv:2203.07622}}.
\bibitem[{Athron et~al.(2024)Athron, Crivellin, Gonzalo, Iguro and
  Sierra}]{Athron:2024rir}
\bibinfo{author}{Athron, P.}, \bibinfo{author}{Crivellin, A.},
  \bibinfo{author}{Gonzalo, T.E.}, \bibinfo{author}{Iguro, S.},
  \bibinfo{author}{Sierra, C.}, \bibinfo{year}{2024}.
\newblock \bibinfo{title}{{Global fit to the 2HDM with generic sources of
  flavour violation using {GAMBIT}}}
  \href{http://arxiv.org/abs/2410.10493}{{\tt arXiv:2410.10493}}.
\bibitem[{{ATLAS collaboration}(2022a)}]{ATLAS:2022vkf}
\bibinfo{author}{{ATLAS collaboration}}, \bibinfo{year}{2022}a.
\newblock \bibinfo{title}{{A detailed map of Higgs boson interactions by the
  ATLAS experiment ten years after the discovery}}.
\newblock \bibinfo{journal}{Nature} \bibinfo{volume}{607},
  \bibinfo{pages}{52--59}.
\newblock \DOIprefix\doi{10.1038/s41586-022-04893-w},
  \href{http://arxiv.org/abs/2207.00092}{{\tt arXiv:2207.00092}}.
  \bibinfo{note}{[Erratum: Nature 612, E24 (2022)]}.
\bibitem[{{ATLAS collaboration}(2022b)}]{ATL-PHYS-PUB-2022-018}
\bibinfo{author}{{ATLAS collaboration}}, \bibinfo{year}{2022}b.
\newblock \bibinfo{title}{{Snowmass White Paper Contribution: Physics with the
  Phase-2 ATLAS and CMS Detectors}}.
\newblock \bibinfo{type}{Technical Report}. CERN. \bibinfo{address}{Geneva}.
\newblock \URLprefix \url{https://cds.cern.ch/record/2805993}.
  \bibinfo{note}{all figures including auxiliary figures are available at
  https://atlas.web.cern.ch/Atlas/GROUPS/PHYSICS/PUBNOTES/ATL-PHYS-PUB-2022-018}.
\bibitem[{Babu(1988)}]{Babu:1988ki}
\bibinfo{author}{Babu, K.S.}, \bibinfo{year}{1988}.
\newblock \bibinfo{title}{{Model of 'Calculable' Majorana Neutrino Masses}}.
\newblock \bibinfo{journal}{Phys. Lett. B} \bibinfo{volume}{203},
  \bibinfo{pages}{132--136}.
\newblock \DOIprefix\doi{10.1016/0370-2693(88)91584-5}.
\bibitem[{Babu and Jana(2017)}]{Babu:2016rcr}
\bibinfo{author}{Babu, K.S.}, \bibinfo{author}{Jana, S.}, \bibinfo{year}{2017}.
\newblock \bibinfo{title}{{Probing Doubly Charged Higgs Bosons at the LHC
  through Photon Initiated Processes}}.
\newblock \bibinfo{journal}{Phys. Rev. D} \bibinfo{volume}{95},
  \bibinfo{pages}{055020}.
\newblock \DOIprefix\doi{10.1103/PhysRevD.95.055020},
  \href{http://arxiv.org/abs/1612.09224}{{\tt arXiv:1612.09224}}.
\bibitem[{Baer et~al.(2013)}]{ILC:2013jhg}
\bibinfo{author}{Baer, H.}, et~al. (\bibinfo{collaboration}{ILC}),
  \bibinfo{year}{2013}.
\newblock \bibinfo{title}{{The International Linear Collider Technical Design
  Report - Volume 2: Physics}}.
\newblock \href{http://arxiv.org/abs/1306.6352}{{\tt arXiv:1306.6352}}.
\bibitem[{Bambade et~al.(2019)}]{Bambade:2019fyw}
\bibinfo{author}{Bambade, P.}, et~al., \bibinfo{year}{2019}.
\newblock \bibinfo{title}{{The International Linear Collider: A Global
  Project}} \href{http://arxiv.org/abs/1903.01629}{{\tt arXiv:1903.01629}}.
\bibitem[{Belloni et~al.(2022)}]{Belloni:2022due}
\bibinfo{author}{Belloni, A.}, et~al., \bibinfo{year}{2022}.
\newblock \bibinfo{title}{{Report of the Topical Group on Electroweak Precision
  Physics and Constraining New Physics for Snowmass 2021}}
  \href{http://arxiv.org/abs/2209.08078}{{\tt arXiv:2209.08078}}.
\bibitem[{Bennett et~al.(2006)}]{Muong-2:2006rrc}
\bibinfo{author}{Bennett, G.W.}, et~al. (\bibinfo{collaboration}{Muon g-2}),
  \bibinfo{year}{2006}.
\newblock \bibinfo{title}{{Final Report of the Muon E821 Anomalous Magnetic
  Moment Measurement at BNL}}.
\newblock \bibinfo{journal}{Phys. Rev. D} \bibinfo{volume}{73},
  \bibinfo{pages}{072003}.
\newblock \DOIprefix\doi{10.1103/PhysRevD.73.072003},
  \href{http://arxiv.org/abs/hep-ex/0602035}{{\tt arXiv:hep-ex/0602035}}.
\bibitem[{Bergstrom and Hulth(1985)}]{Bergstrom:1985hp}
\bibinfo{author}{Bergstrom, L.}, \bibinfo{author}{Hulth, G.},
  \bibinfo{year}{1985}.
\newblock \bibinfo{title}{{Induced Higgs Couplings to Neutral Bosons in $e^+
  e^-$ Collisions}}.
\newblock \bibinfo{journal}{Nucl. Phys. B} \bibinfo{volume}{259},
  \bibinfo{pages}{137--155}.
\newblock \DOIprefix\doi{10.1016/0550-3213(85)90302-5}.
  \bibinfo{note}{[Erratum: Nucl.Phys.B 276, 744--744 (1986)]}.
\bibitem[{Bernardi et~al.(2022)}]{Bernardi:2022hny}
\bibinfo{author}{Bernardi, G.}, et~al., \bibinfo{year}{2022}.
\newblock \bibinfo{title}{{The Future Circular Collider: a Summary for the US
  2021 Snowmass Process}} \href{http://arxiv.org/abs/2203.06520}{{\tt
  arXiv:2203.06520}}.
\bibitem[{Bigaran et~al.(2023)Bigaran, He, Schmidt, Valencia and
  Volkas}]{Bigaran:2022giz}
\bibinfo{author}{Bigaran, I.}, \bibinfo{author}{He, X.G.},
  \bibinfo{author}{Schmidt, M.A.}, \bibinfo{author}{Valencia, G.},
  \bibinfo{author}{Volkas, R.}, \bibinfo{year}{2023}.
\newblock \bibinfo{title}{{Lepton-flavor-violating tau decays from triality}}.
\newblock \bibinfo{journal}{Phys. Rev. D} \bibinfo{volume}{107},
  \bibinfo{pages}{055001}.
\newblock \DOIprefix\doi{10.1103/PhysRevD.107.055001},
  \href{http://arxiv.org/abs/2212.09760}{{\tt arXiv:2212.09760}}.
\bibitem[{de~Blas et~al.(2015)de~Blas, Chala, Perez-Victoria and
  Santiago}]{deBlas:2014mba}
\bibinfo{author}{de~Blas, J.}, \bibinfo{author}{Chala, M.},
  \bibinfo{author}{Perez-Victoria, M.}, \bibinfo{author}{Santiago, J.},
  \bibinfo{year}{2015}.
\newblock \bibinfo{title}{{Observable Effects of General New Scalar
  Particles}}.
\newblock \bibinfo{journal}{JHEP} \bibinfo{volume}{04}, \bibinfo{pages}{078}.
\newblock \DOIprefix\doi{10.1007/JHEP04(2015)078},
  \href{http://arxiv.org/abs/1412.8480}{{\tt arXiv:1412.8480}}.
\bibitem[{de~Blas et~al.(2020)}]{deBlas:2019rxi}
\bibinfo{author}{de~Blas, J.}, et~al., \bibinfo{year}{2020}.
\newblock \bibinfo{title}{{Higgs Boson Studies at Future Particle Colliders}}.
\newblock \bibinfo{journal}{JHEP} \bibinfo{volume}{01}, \bibinfo{pages}{139}.
\newblock \DOIprefix\doi{10.1007/JHEP01(2020)139},
  \href{http://arxiv.org/abs/1905.03764}{{\tt arXiv:1905.03764}}.
\bibitem[{Boccaletti et~al.(2024)}]{Boccaletti:2024guq}
\bibinfo{author}{Boccaletti, A.}, et~al., \bibinfo{year}{2024}.
\newblock \bibinfo{title}{{High precision calculation of the hadronic vacuum
  polarisation contribution to the muon anomaly}}
  \href{http://arxiv.org/abs/2407.10913}{{\tt arXiv:2407.10913}}.
\bibitem[{Borsanyi et~al.(2021)}]{Borsanyi:2020mff}
\bibinfo{author}{Borsanyi, S.}, et~al., \bibinfo{year}{2021}.
\newblock \bibinfo{title}{{Leading hadronic contribution to the muon magnetic
  moment from lattice QCD}}.
\newblock \bibinfo{journal}{Nature} \bibinfo{volume}{593},
  \bibinfo{pages}{51--55}.
\newblock \DOIprefix\doi{10.1038/s41586-021-03418-1},
  \href{http://arxiv.org/abs/2002.12347}{{\tt arXiv:2002.12347}}.
\bibitem[{Boto et~al.(2024)Boto, Das, Romao, Saha and Silva}]{Boto:2023bpg}
\bibinfo{author}{Boto, R.}, \bibinfo{author}{Das, D.}, \bibinfo{author}{Romao,
  J.C.}, \bibinfo{author}{Saha, I.}, \bibinfo{author}{Silva, J.P.},
  \bibinfo{year}{2024}.
\newblock \bibinfo{title}{{New physics interpretations for nonstandard values
  of h\textrightarrow{}Z\ensuremath{\gamma}}}.
\newblock \bibinfo{journal}{Phys. Rev. D} \bibinfo{volume}{109},
  \bibinfo{pages}{095002}.
\newblock \DOIprefix\doi{10.1103/PhysRevD.109.095002},
  \href{http://arxiv.org/abs/2312.13050}{{\tt arXiv:2312.13050}}.
\bibitem[{Cao et~al.(2011)Cao, Damanik, Ma and Wegman}]{Cao:2011df}
\bibinfo{author}{Cao, Q.H.}, \bibinfo{author}{Damanik, A.},
  \bibinfo{author}{Ma, E.}, \bibinfo{author}{Wegman, D.}, \bibinfo{year}{2011}.
\newblock \bibinfo{title}{{Probing Lepton Flavor Triality with Higgs Boson
  Decay}}.
\newblock \bibinfo{journal}{Phys. Rev. D} \bibinfo{volume}{83},
  \bibinfo{pages}{093012}.
\newblock \DOIprefix\doi{10.1103/PhysRevD.83.093012},
  \href{http://arxiv.org/abs/1103.0008}{{\tt arXiv:1103.0008}}.
\bibitem[{Cepeda et~al.(2019)}]{Cepeda:2019klc}
\bibinfo{author}{Cepeda, M.}, et~al., \bibinfo{year}{2019}.
\newblock \bibinfo{title}{{Report from Working Group 2}: {Higgs Physics at the
  HL-LHC and HE-LHC}}.
\newblock \bibinfo{journal}{CERN Yellow Rep. Monogr.} \bibinfo{volume}{7},
  \bibinfo{pages}{221--584}.
\newblock \DOIprefix\doi{10.23731/CYRM-2019-007.221},
  \href{http://arxiv.org/abs/1902.00134}{{\tt arXiv:1902.00134}}.
\bibitem[{Cheng et~al.(2022)}]{CEPCPhysicsStudyGroup:2022uwl}
\bibinfo{author}{Cheng, H.}, et~al. (\bibinfo{collaboration}{CEPC Physics Study
  Group}), \bibinfo{year}{2022}.
\newblock \bibinfo{title}{{The Physics potential of the CEPC. Prepared for the
  US Snowmass Community Planning Exercise (Snowmass 2021)}}, in:
  \bibinfo{booktitle}{{Snowmass 2021}}.
\newblock \href{http://arxiv.org/abs/2205.08553}{{\tt arXiv:2205.08553}}.
\bibitem[{Christensen et~al.(2011)Christensen, de~Aquino, Degrande, Duhr, Fuks,
  Herquet, Maltoni and Schumann}]{Christensen:2009jx}
\bibinfo{author}{Christensen, N.D.}, \bibinfo{author}{de~Aquino, P.},
  \bibinfo{author}{Degrande, C.}, \bibinfo{author}{Duhr, C.},
  \bibinfo{author}{Fuks, B.}, \bibinfo{author}{Herquet, M.},
  \bibinfo{author}{Maltoni, F.}, \bibinfo{author}{Schumann, S.},
  \bibinfo{year}{2011}.
\newblock \bibinfo{title}{{A Comprehensive approach to new physics
  simulations}}.
\newblock \bibinfo{journal}{Eur. Phys. J.} \bibinfo{volume}{C71},
  \bibinfo{pages}{1541}.
\newblock \DOIprefix\doi{10.1140/epjc/s10052-011-1541-5},
  \href{http://arxiv.org/abs/0906.2474}{{\tt arXiv:0906.2474}}.
\bibitem[{Crivellin et~al.(2019)Crivellin, Ghezzi, Panizzi, Pruna and
  Signer}]{Crivellin:2018ahj}
\bibinfo{author}{Crivellin, A.}, \bibinfo{author}{Ghezzi, M.},
  \bibinfo{author}{Panizzi, L.}, \bibinfo{author}{Pruna, G.M.},
  \bibinfo{author}{Signer, A.}, \bibinfo{year}{2019}.
\newblock \bibinfo{title}{{Low- and high-energy phenomenology of a doubly
  charged scalar}}.
\newblock \bibinfo{journal}{Phys. Rev. D} \bibinfo{volume}{99},
  \bibinfo{pages}{035004}.
\newblock \DOIprefix\doi{10.1103/PhysRevD.99.035004},
  \href{http://arxiv.org/abs/1807.10224}{{\tt arXiv:1807.10224}}.
\bibitem[{Dainese et~al.(2019)Dainese, Mangano, Meyer, Nisati, Salam and
  Vesterinen}]{Dainese:2019rgk}
\bibinfo{editor}{Dainese, A.}, \bibinfo{editor}{Mangano, M.},
  \bibinfo{editor}{Meyer, A.B.}, \bibinfo{editor}{Nisati, A.},
  \bibinfo{editor}{Salam, G.}, \bibinfo{editor}{Vesterinen, M.A.} (Eds.),
  \bibinfo{year}{2019}.
\newblock \bibinfo{title}{{Report on the Physics at the HL-LHC,and Perspectives
  for the HE-LHC}}. volume \bibinfo{volume}{7/2019} of
  \textit{\bibinfo{series}{CERN Yellow Reports: Monographs}}.
\newblock \bibinfo{publisher}{CERN}, \bibinfo{address}{Geneva, Switzerland}.
\newblock \DOIprefix\doi{10.23731/CYRM-2019-007}.
\bibitem[{Das and Santamaria(2016)}]{Das:2016bir}
\bibinfo{author}{Das, D.}, \bibinfo{author}{Santamaria, A.},
  \bibinfo{year}{2016}.
\newblock \bibinfo{title}{{Updated scalar sector constraints in the Higgs
  triplet model}}.
\newblock \bibinfo{journal}{Phys. Rev. D} \bibinfo{volume}{94},
  \bibinfo{pages}{015015}.
\newblock \DOIprefix\doi{10.1103/PhysRevD.94.015015},
  \href{http://arxiv.org/abs/1604.08099}{{\tt arXiv:1604.08099}}.
\bibitem[{Dawson and Valencia(1995)}]{Dawson:1994fa}
\bibinfo{author}{Dawson, S.}, \bibinfo{author}{Valencia, G.},
  \bibinfo{year}{1995}.
\newblock \bibinfo{title}{{Bounds on anomalous gauge boson couplings from
  partial Z widths at LEP}}.
\newblock \bibinfo{journal}{Nucl. Phys. B} \bibinfo{volume}{439},
  \bibinfo{pages}{3--22}.
\newblock \DOIprefix\doi{10.1016/0550-3213(95)00042-Q},
  \href{http://arxiv.org/abs/hep-ph/9410364}{{\tt arXiv:hep-ph/9410364}}.
\bibitem[{Dawson et~al.(2022)}]{Dawson:2022zbb}
\bibinfo{author}{Dawson, S.}, et~al., \bibinfo{year}{2022}.
\newblock \bibinfo{title}{{Report of the Topical Group on Higgs Physics for
  Snowmass 2021: The Case for Precision Higgs Physics}}, in:
  \bibinfo{booktitle}{{Snowmass 2021}}.
\newblock \href{http://arxiv.org/abs/2209.07510}{{\tt arXiv:2209.07510}}.
\bibitem[{Dong et~al.(2018)}]{CEPCStudyGroup:2018ghi}
\bibinfo{author}{Dong, M.}, et~al. (\bibinfo{collaboration}{CEPC Study Group}),
  \bibinfo{year}{2018}.
\newblock \bibinfo{title}{{CEPC Conceptual Design Report: Volume 2 - Physics \&
  Detector}} \href{http://arxiv.org/abs/1811.10545}{{\tt arXiv:1811.10545}}.
\bibitem[{de~Florian et~al.(2016)}]{LHCHiggsCrossSectionWorkingGroup:2016ypw}
\bibinfo{author}{de~Florian, D.}, et~al. (\bibinfo{collaboration}{LHC Higgs
  Cross Section Working Group}), \bibinfo{year}{2016}.
\newblock \bibinfo{title}{{Handbook of LHC Higgs Cross Sections: 4. Deciphering
  the Nature of the Higgs Sector}} \bibinfo{volume}{2/2017}.
\newblock \DOIprefix\doi{10.23731/CYRM-2017-002},
  \href{http://arxiv.org/abs/1610.07922}{{\tt arXiv:1610.07922}}.
\bibitem[{Fuks et~al.(2020)Fuks, Nemev\v{s}ek and Ruiz}]{Fuks:2019clu}
\bibinfo{author}{Fuks, B.}, \bibinfo{author}{Nemev\v{s}ek, M.},
  \bibinfo{author}{Ruiz, R.}, \bibinfo{year}{2020}.
\newblock \bibinfo{title}{{Doubly Charged Higgs Boson Production at Hadron
  Colliders}}.
\newblock \bibinfo{journal}{Phys. Rev. D} \bibinfo{volume}{101},
  \bibinfo{pages}{075022}.
\newblock \DOIprefix\doi{10.1103/PhysRevD.101.075022},
  \href{http://arxiv.org/abs/1912.08975}{{\tt arXiv:1912.08975}}.
\bibitem[{Gao(2022)}]{Gao:2022lew}
\bibinfo{author}{Gao, J.} (\bibinfo{collaboration}{CEPC Accelerator Study
  Group}), \bibinfo{year}{2022}.
\newblock \bibinfo{title}{{Snowmass2021 White Paper AF3-CEPC}}
  \href{http://arxiv.org/abs/2203.09451}{{\tt arXiv:2203.09451}}.
\bibitem[{Gunion et~al.(2000)Gunion, Haber, Kane and Dawson}]{Gunion:1989we}
\bibinfo{author}{Gunion, J.F.}, \bibinfo{author}{Haber, H.E.},
  \bibinfo{author}{Kane, G.L.}, \bibinfo{author}{Dawson, S.},
  \bibinfo{year}{2000}.
\newblock \bibinfo{title}{{The Higgs Hunter's Guide}}.
  volume~\bibinfo{volume}{80}.
\bibitem[{Hahn(2001)}]{Hahn:2000kx}
\bibinfo{author}{Hahn, T.}, \bibinfo{year}{2001}.
\newblock \bibinfo{title}{{Generating Feynman diagrams and amplitudes with
  FeynArts 3}}.
\newblock \bibinfo{journal}{Comput. Phys. Commun.} \bibinfo{volume}{140},
  \bibinfo{pages}{418--431}.
\newblock \DOIprefix\doi{10.1016/S0010-4655(01)00290-9},
  \href{http://arxiv.org/abs/hep-ph/0012260}{{\tt arXiv:hep-ph/0012260}}.
\bibitem[{Hahn and Perez-Victoria(1999)}]{Hahn:1998yk}
\bibinfo{author}{Hahn, T.}, \bibinfo{author}{Perez-Victoria, M.},
  \bibinfo{year}{1999}.
\newblock \bibinfo{title}{{Automatized one loop calculations in four-dimensions
  and D-dimensions}}.
\newblock \bibinfo{journal}{Comput. Phys. Commun.} \bibinfo{volume}{118},
  \bibinfo{pages}{153--165}.
\newblock \DOIprefix\doi{10.1016/S0010-4655(98)00173-8},
  \href{http://arxiv.org/abs/hep-ph/9807565}{{\tt arXiv:hep-ph/9807565}}.
\bibitem[{Hamada et~al.(2022)Hamada, Kitano, Matsudo, Takaura and
  Yoshida}]{Hamada:2022mua}
\bibinfo{author}{Hamada, Y.}, \bibinfo{author}{Kitano, R.},
  \bibinfo{author}{Matsudo, R.}, \bibinfo{author}{Takaura, H.},
  \bibinfo{author}{Yoshida, M.}, \bibinfo{year}{2022}.
\newblock \bibinfo{title}{{$\mu$TRISTAN}}.
\newblock \bibinfo{journal}{PTEP} \bibinfo{volume}{2022},
  \bibinfo{pages}{053B02}.
\newblock \DOIprefix\doi{10.1093/ptep/ptac059},
  \href{http://arxiv.org/abs/2201.06664}{{\tt arXiv:2201.06664}}.
\bibitem[{He et~al.(2024)He, Huang, Li and Liu}]{He:2024bxi}
\bibinfo{author}{He, X.G.}, \bibinfo{author}{Huang, Z.L.}, \bibinfo{author}{Li,
  M.W.}, \bibinfo{author}{Liu, C.W.}, \bibinfo{year}{2024}.
\newblock \bibinfo{title}{{The SM expected branching ratio for h
  \textrightarrow{} \ensuremath{\gamma}\ensuremath{\gamma} and an excess for h
  \textrightarrow{} Z\ensuremath{\gamma}}}.
\newblock \bibinfo{journal}{JHEP} \bibinfo{volume}{10}, \bibinfo{pages}{135}.
\newblock \DOIprefix\doi{10.1007/JHEP10(2024)135},
  \href{http://arxiv.org/abs/2402.08190}{{\tt arXiv:2402.08190}}.
\bibitem[{He et~al.(2006)He, Keum and Volkas}]{He:2006dk}
\bibinfo{author}{He, X.G.}, \bibinfo{author}{Keum, Y.Y.},
  \bibinfo{author}{Volkas, R.R.}, \bibinfo{year}{2006}.
\newblock \bibinfo{title}{{A(4) flavor symmetry breaking scheme for
  understanding quark and neutrino mixing angles}}.
\newblock \bibinfo{journal}{JHEP} \bibinfo{volume}{04}, \bibinfo{pages}{039}.
\newblock \DOIprefix\doi{10.1088/1126-6708/2006/04/039},
  \href{http://arxiv.org/abs/hep-ph/0601001}{{\tt arXiv:hep-ph/0601001}}.
\bibitem[{Holthausen et~al.(2013)Holthausen, Lindner and
  Schmidt}]{Holthausen:2012wz}
\bibinfo{author}{Holthausen, M.}, \bibinfo{author}{Lindner, M.},
  \bibinfo{author}{Schmidt, M.A.}, \bibinfo{year}{2013}.
\newblock \bibinfo{title}{{Lepton flavor at the electroweak scale: A complete
  $A_{4}$ model}}.
\newblock \bibinfo{journal}{Phys. Rev. D} \bibinfo{volume}{87},
  \bibinfo{pages}{033006}.
\newblock \DOIprefix\doi{10.1103/PhysRevD.87.033006},
  \href{http://arxiv.org/abs/1211.5143}{{\tt arXiv:1211.5143}}.
\bibitem[{Hong et~al.(2024)Hong, Le, Phuong, Hoi, Ngan and Nha}]{Hong:2023mwr}
\bibinfo{author}{Hong, T.T.}, \bibinfo{author}{Le, V.K.},
  \bibinfo{author}{Phuong, L.T.T.}, \bibinfo{author}{Hoi, N..C.},
  \bibinfo{author}{Ngan, N.T.K.}, \bibinfo{author}{Nha, N.H.T.},
  \bibinfo{year}{2024}.
\newblock \bibinfo{title}{{Decays of Standard Model\textendash{}Like Higgs
  Boson $h \rightarrow\gamma\gamma, Z \gamma$in a Minimal Left-Right Symmetric
  Model}}.
\newblock \bibinfo{journal}{PTEP} \bibinfo{volume}{2024},
  \bibinfo{pages}{033B04}.
\newblock \DOIprefix\doi{10.1093/ptep/ptae029},
  \href{http://arxiv.org/abs/2312.11045}{{\tt arXiv:2312.11045}}.
  \bibinfo{note}{[Erratum: PTEP 2024, 059201 (2024)]}.
\bibitem[{Kannike(2012)}]{Kannike:2012pe}
\bibinfo{author}{Kannike, K.}, \bibinfo{year}{2012}.
\newblock \bibinfo{title}{{Vacuum Stability Conditions From Copositivity
  Criteria}}.
\newblock \bibinfo{journal}{Eur. Phys. J. C} \bibinfo{volume}{72},
  \bibinfo{pages}{2093}.
\newblock \DOIprefix\doi{10.1140/epjc/s10052-012-2093-z},
  \href{http://arxiv.org/abs/1205.3781}{{\tt arXiv:1205.3781}}.
\bibitem[{Lavoura(2003)}]{Lavoura:2003xp}
\bibinfo{author}{Lavoura, L.}, \bibinfo{year}{2003}.
\newblock \bibinfo{title}{{General formulae for f(1) ---\ensuremath{>} f(2)
  gamma}}.
\newblock \bibinfo{journal}{Eur. Phys. J. C} \bibinfo{volume}{29},
  \bibinfo{pages}{191--195}.
\newblock \DOIprefix\doi{10.1140/epjc/s2003-01212-7},
  \href{http://arxiv.org/abs/hep-ph/0302221}{{\tt arXiv:hep-ph/0302221}}.
\bibitem[{Li and Schmidt(2019a)}]{Li:2019xvv}
\bibinfo{author}{Li, T.}, \bibinfo{author}{Schmidt, M.A.},
  \bibinfo{year}{2019}a.
\newblock \bibinfo{title}{{Sensitivity of future lepton colliders and
  low-energy experiments to charged lepton flavor violation from bileptons}}.
\newblock \bibinfo{journal}{Phys. Rev. D} \bibinfo{volume}{100},
  \bibinfo{pages}{115007}.
\newblock \DOIprefix\doi{10.1103/PhysRevD.100.115007},
  \href{http://arxiv.org/abs/1907.06963}{{\tt arXiv:1907.06963}}.
\bibitem[{Li and Schmidt(2019b)}]{Li:2018cod}
\bibinfo{author}{Li, T.}, \bibinfo{author}{Schmidt, M.A.},
  \bibinfo{year}{2019}b.
\newblock \bibinfo{title}{{Sensitivity of future lepton colliders to the search
  for charged lepton flavor violation}}.
\newblock \bibinfo{journal}{Phys. Rev. D} \bibinfo{volume}{99},
  \bibinfo{pages}{055038}.
\newblock \DOIprefix\doi{10.1103/PhysRevD.99.055038},
  \href{http://arxiv.org/abs/1809.07924}{{\tt arXiv:1809.07924}}.
\bibitem[{Lichtenstein et~al.(2023)Lichtenstein, Schmidt, Valencia and
  Volkas}]{Lichtenstein:2023iut}
\bibinfo{author}{Lichtenstein, G.}, \bibinfo{author}{Schmidt, M.A.},
  \bibinfo{author}{Valencia, G.}, \bibinfo{author}{Volkas, R.R.},
  \bibinfo{year}{2023}.
\newblock \bibinfo{title}{{Complementarity of \ensuremath{\mu}TRISTAN and Belle
  II in searches for charged-lepton flavour violation}}.
\newblock \bibinfo{journal}{Phys. Lett. B} \bibinfo{volume}{845},
  \bibinfo{pages}{138144}.
\newblock \DOIprefix\doi{10.1016/j.physletb.2023.138144},
  \href{http://arxiv.org/abs/2307.11369}{{\tt arXiv:2307.11369}}.
\bibitem[{de~Melo et~al.(2019)de~Melo, Queiroz and Villamizar}]{deMelo:2019asm}
\bibinfo{author}{de~Melo, T.B.}, \bibinfo{author}{Queiroz, F.S.},
  \bibinfo{author}{Villamizar, Y.}, \bibinfo{year}{2019}.
\newblock \bibinfo{title}{{Doubly Charged Scalar at the High-Luminosity and
  High-Energy LHC}}.
\newblock \bibinfo{journal}{Int. J. Mod. Phys. A} \bibinfo{volume}{34},
  \bibinfo{pages}{1950157}.
\newblock \DOIprefix\doi{10.1142/S0217751X19501574},
  \href{http://arxiv.org/abs/1909.07429}{{\tt arXiv:1909.07429}}.
\bibitem[{Mertig et~al.(1991)Mertig, Bohm and Denner}]{Mertig:1990an}
\bibinfo{author}{Mertig, R.}, \bibinfo{author}{Bohm, M.},
  \bibinfo{author}{Denner, A.}, \bibinfo{year}{1991}.
\newblock \bibinfo{title}{{FEYN CALC: Computer algebraic calculation of Feynman
  amplitudes}}.
\newblock \bibinfo{journal}{Comput. Phys. Commun.} \bibinfo{volume}{64},
  \bibinfo{pages}{345--359}.
\newblock \DOIprefix\doi{10.1016/0010-4655(91)90130-D}.
\bibitem[{Mlynarikova(2023)}]{Mlynarikova:2023bvx}
\bibinfo{author}{Mlynarikova, M.} (\bibinfo{collaboration}{ATLAS, CMS}),
  \bibinfo{year}{2023}.
\newblock \bibinfo{title}{{Higgs Physics at HL-LHC}}, in:
  \bibinfo{booktitle}{{30th International Workshop on Deep-Inelastic Scattering
  and Related Subjects}}.
\newblock \href{http://arxiv.org/abs/2307.07772}{{\tt arXiv:2307.07772}}.
\bibitem[{Morel et~al.(2020)Morel, Yao, Clad\'e and
  Guellati-Kh\'elifa}]{Morel:2020dww}
\bibinfo{author}{Morel, L.}, \bibinfo{author}{Yao, Z.},
  \bibinfo{author}{Clad\'e, P.}, \bibinfo{author}{Guellati-Kh\'elifa, S.},
  \bibinfo{year}{2020}.
\newblock \bibinfo{title}{{Determination of the fine-structure constant with an
  accuracy of 81 parts per trillion}}.
\newblock \bibinfo{journal}{Nature} \bibinfo{volume}{588},
  \bibinfo{pages}{61--65}.
\newblock \DOIprefix\doi{10.1038/s41586-020-2964-7}.
\bibitem[{Muramatsu et~al.(2016)Muramatsu, Nomura and
  Shimizu}]{Muramatsu:2016bda}
\bibinfo{author}{Muramatsu, Y.}, \bibinfo{author}{Nomura, T.},
  \bibinfo{author}{Shimizu, Y.}, \bibinfo{year}{2016}.
\newblock \bibinfo{title}{{Mass limit for light flavon with residual Z$_{3}$
  symmetry}}.
\newblock \bibinfo{journal}{JHEP} \bibinfo{volume}{03}, \bibinfo{pages}{192}.
\newblock \DOIprefix\doi{10.1007/JHEP03(2016)192},
  \href{http://arxiv.org/abs/1601.04788}{{\tt arXiv:1601.04788}}.
\bibitem[{Nowakowski et~al.(2005)Nowakowski, Paschos and
  Rodriguez}]{Nowakowski:2004cv}
\bibinfo{author}{Nowakowski, M.}, \bibinfo{author}{Paschos, E.A.},
  \bibinfo{author}{Rodriguez, J.M.}, \bibinfo{year}{2005}.
\newblock \bibinfo{title}{{All electromagnetic form-factors}}.
\newblock \bibinfo{journal}{Eur. J. Phys.} \bibinfo{volume}{26},
  \bibinfo{pages}{545--560}.
\newblock \DOIprefix\doi{10.1088/0143-0807/26/4/001},
  \href{http://arxiv.org/abs/physics/0402058}{{\tt arXiv:physics/0402058}}.
\bibitem[{Parker et~al.(2018)Parker, Yu, Zhong, Estey and
  M\"uller}]{Parker:2018vye}
\bibinfo{author}{Parker, R.H.}, \bibinfo{author}{Yu, C.},
  \bibinfo{author}{Zhong, W.}, \bibinfo{author}{Estey, B.},
  \bibinfo{author}{M\"uller, H.}, \bibinfo{year}{2018}.
\newblock \bibinfo{title}{{Measurement of the fine-structure constant as a test
  of the Standard Model}}.
\newblock \bibinfo{journal}{Science} \bibinfo{volume}{360},
  \bibinfo{pages}{191}.
\newblock \DOIprefix\doi{10.1126/science.aap7706},
  \href{http://arxiv.org/abs/1812.04130}{{\tt arXiv:1812.04130}}.
\bibitem[{Pascoli and Zhou(2016)}]{Pascoli:2016wlt}
\bibinfo{author}{Pascoli, S.}, \bibinfo{author}{Zhou, Y.L.},
  \bibinfo{year}{2016}.
\newblock \bibinfo{title}{{Flavon-induced connections between lepton flavour
  mixing and charged lepton flavour violation processes}}.
\newblock \bibinfo{journal}{JHEP} \bibinfo{volume}{10}, \bibinfo{pages}{145}.
\newblock \DOIprefix\doi{10.1007/JHEP10(2016)145},
  \href{http://arxiv.org/abs/1607.05599}{{\tt arXiv:1607.05599}}.
\bibitem[{Patel(2015)}]{Patel:2015tea}
\bibinfo{author}{Patel, H.H.}, \bibinfo{year}{2015}.
\newblock \bibinfo{title}{{Package-X: A Mathematica package for the analytic
  calculation of one-loop integrals}}.
\newblock \bibinfo{journal}{Comput. Phys. Commun.} \bibinfo{volume}{197},
  \bibinfo{pages}{276--290}.
\newblock \DOIprefix\doi{10.1016/j.cpc.2015.08.017},
  \href{http://arxiv.org/abs/1503.01469}{{\tt arXiv:1503.01469}}.
\bibitem[{Patel(2017)}]{Patel:2016fam}
\bibinfo{author}{Patel, H.H.}, \bibinfo{year}{2017}.
\newblock \bibinfo{title}{{Package-X 2.0: A Mathematica package for the
  analytic calculation of one-loop integrals}}.
\newblock \bibinfo{journal}{Comput. Phys. Commun.} \bibinfo{volume}{218},
  \bibinfo{pages}{66--70}.
\newblock \DOIprefix\doi{10.1016/j.cpc.2017.04.015},
  \href{http://arxiv.org/abs/1612.00009}{{\tt arXiv:1612.00009}}.
\bibitem[{Ruiz(2022)}]{Ruiz:2022sct}
\bibinfo{author}{Ruiz, R.}, \bibinfo{year}{2022}.
\newblock \bibinfo{title}{{Doubly charged Higgs boson production at hadron
  colliders II: a Zee-Babu case study}}.
\newblock \bibinfo{journal}{JHEP} \bibinfo{volume}{10}, \bibinfo{pages}{200}.
\newblock \DOIprefix\doi{10.1007/JHEP10(2022)200},
  \href{http://arxiv.org/abs/2206.14833}{{\tt arXiv:2206.14833}}.
\bibitem[{Schael et~al.(2006)}]{ALEPH:2005ab}
\bibinfo{author}{Schael, S.}, et~al. (\bibinfo{collaboration}{ALEPH, DELPHI,
  L3, OPAL, SLD, LEP Electroweak Working Group, SLD Electroweak Group, SLD
  Heavy Flavour Group}), \bibinfo{year}{2006}.
\newblock \bibinfo{title}{{Precision electroweak measurements on the $Z$
  resonance}}.
\newblock \bibinfo{journal}{Phys. Rept.} \bibinfo{volume}{427},
  \bibinfo{pages}{257--454}.
\newblock \DOIprefix\doi{10.1016/j.physrep.2005.12.006},
  \href{http://arxiv.org/abs/hep-ex/0509008}{{\tt arXiv:hep-ex/0509008}}.
\bibitem[{Schwartz(2014)}]{Schwartz:2014sze}
\bibinfo{author}{Schwartz, M.D.}, \bibinfo{year}{2014}.
\newblock \bibinfo{title}{{Quantum Field Theory and the Standard Model}}.
\newblock \bibinfo{publisher}{Cambridge University Press}.
\bibitem[{Shtabovenko(2017)}]{Shtabovenko:2016whf}
\bibinfo{author}{Shtabovenko, V.}, \bibinfo{year}{2017}.
\newblock \bibinfo{title}{{FeynHelpers: Connecting FeynCalc to FIRE and
  Package-X}}.
\newblock \bibinfo{journal}{Comput. Phys. Commun.} \bibinfo{volume}{218},
  \bibinfo{pages}{48--65}.
\newblock \DOIprefix\doi{10.1016/j.cpc.2017.04.014},
  \href{http://arxiv.org/abs/1611.06793}{{\tt arXiv:1611.06793}}.
\bibitem[{Shtabovenko et~al.(2016)Shtabovenko, Mertig and
  Orellana}]{Shtabovenko:2016sxi}
\bibinfo{author}{Shtabovenko, V.}, \bibinfo{author}{Mertig, R.},
  \bibinfo{author}{Orellana, F.}, \bibinfo{year}{2016}.
\newblock \bibinfo{title}{{New Developments in FeynCalc 9.0}}.
\newblock \bibinfo{journal}{Comput. Phys. Commun.} \bibinfo{volume}{207},
  \bibinfo{pages}{432--444}.
\newblock \DOIprefix\doi{10.1016/j.cpc.2016.06.008},
  \href{http://arxiv.org/abs/1601.01167}{{\tt arXiv:1601.01167}}.
\bibitem[{Shtabovenko et~al.(2020)Shtabovenko, Mertig and
  Orellana}]{Shtabovenko:2020gxv}
\bibinfo{author}{Shtabovenko, V.}, \bibinfo{author}{Mertig, R.},
  \bibinfo{author}{Orellana, F.}, \bibinfo{year}{2020}.
\newblock \bibinfo{title}{{FeynCalc 9.3: New features and improvements}}.
\newblock \bibinfo{journal}{Comput. Phys. Commun.} \bibinfo{volume}{256},
  \bibinfo{pages}{107478}.
\newblock \DOIprefix\doi{10.1016/j.cpc.2020.107478},
  \href{http://arxiv.org/abs/2001.04407}{{\tt arXiv:2001.04407}}.
\bibitem[{Shtabovenko et~al.(2025)Shtabovenko, Mertig and
  Orellana}]{Shtabovenko:2023idz}
\bibinfo{author}{Shtabovenko, V.}, \bibinfo{author}{Mertig, R.},
  \bibinfo{author}{Orellana, F.}, \bibinfo{year}{2025}.
\newblock \bibinfo{title}{{FeynCalc 10: Do multiloop integrals dream of
  computer codes?}}
\newblock \bibinfo{journal}{Comput. Phys. Commun.} \bibinfo{volume}{306},
  \bibinfo{pages}{109357}.
\newblock \DOIprefix\doi{10.1016/j.cpc.2024.109357},
  \href{http://arxiv.org/abs/2312.14089}{{\tt arXiv:2312.14089}}.
\bibitem[{Xu(2023)}]{Xu:2023ene}
\bibinfo{author}{Xu, F.}, \bibinfo{year}{2023}.
\newblock \bibinfo{title}{{Neutral and doubly charged scalars at future lepton
  colliders}}.
\newblock \bibinfo{journal}{Phys. Rev. D} \bibinfo{volume}{108},
  \bibinfo{pages}{036002}.
\newblock \DOIprefix\doi{10.1103/PhysRevD.108.036002},
  \href{http://arxiv.org/abs/2302.08653}{{\tt arXiv:2302.08653}}.
\bibitem[{Zee(1986)}]{Zee:1985id}
\bibinfo{author}{Zee, A.}, \bibinfo{year}{1986}.
\newblock \bibinfo{title}{{Quantum Numbers of Majorana Neutrino Masses}}.
\newblock \bibinfo{journal}{Nucl. Phys. B} \bibinfo{volume}{264},
  \bibinfo{pages}{99--110}.
\newblock \DOIprefix\doi{10.1016/0550-3213(86)90475-X}.

\end{thebibliography}

\end{document}